\documentclass[aps,pre,reprint,amsmath,amssymb,groupedaddress,superscriptaddress]{revtex4-1}
\usepackage{color}
\usepackage{graphicx}
\usepackage[normalem]{ulem}
\usepackage{comment}
\usepackage{xcolor,cancel}
\usepackage{enumerate}

\newif\ifHighlitedChanges
\def\ifHighlitedChanges{\iftrue}
\ifHighlitedChanges
  
  \def\STRIKE#1{{\color{red}\sout{#1}}}
\else
  
  \def\STRIKE#1{\relax}
\fi

\begin{document}

\bibliographystyle{apsrev}
\title{Electron transfer across a thermal gradient}
\author{Galen T. Craven}
\affiliation{Department of Chemistry, University of Pennsylvania, Philadelphia, PA  19104, USA} 
\author{Abraham Nitzan}
\affiliation{Department of Chemistry, University of Pennsylvania, Philadelphia, PA  19104, USA} 
\affiliation{School of Chemistry, Tel Aviv University, Tel Aviv 69978, Israel}


\begin{abstract}
Charge transfer is a fundamental process that underlies a multitude of 
phenomena in chemistry and biology.
Recent advances in observing and manipulating charge and heat transport at the nanoscale, 
and recently developed techniques for monitoring temperature at high temporal and spatial resolution, 
imply the need for considering electron transfer across thermal gradients. 
Here, a theory is developed  
for the rate of electron transfer and the associated heat transport between
donor-acceptor pairs located at sites of different temperatures.
To this end, through application of a generalized multidimensional transition state theory, the traditional Arrhenius picture of activation energy 
as a single point on a free energy surface is replaced with a bithermal property that is derived from
statistical weighting over all configurations 
where the reactant and product states are equienergetic.
The flow of energy associated with the electron transfer process is also examined, leading to relations between the rate of heat exchange among the donor and acceptor sites as functions of the temperature difference and the electronic driving bias. In particular, we find that an open electron transfer channel contributes to enhanced heat transport between sites even when they are in electronic equilibrium. The presented results provide a unified theory for charge transport and the associated heat conduction between sites at different temperatures. 
\end{abstract}
\maketitle
\section{Introduction}
The study of electronic transport in molecular nano-junctions naturally involves 
consideration of inelastic transport, where the transporting electron can exchange energy with 
underlying nuclear motions \cite{Nitzan2007jpcm,Dubi2011}. Such studies have been motivated by the use of inelastic
tunneling spectroscopy, and more recently Raman spectroscopy, as diagnostic tools on one hand, 
and by considerations of junction stability on the other.  In parallel, there has been an increasing 
interest in vibrational heat transport in nanostructures and their interfaces with bulk substrates \cite{Cahill2002,Cahill2003,Leitner2008,Leitner2015,Li2012,Dhar2008,Luo2013,Rubtsova2015acr,Rubtsova2015}
focusing on structure-transport correlations \cite{Nitzan2003thermal,Mensah2004,Marconnet2013,AlGhalith2016},  
molecule-substrate coupling \cite{Chen2005,Losego2012,OBrien2013}, ballistic and 
diffusive transport processes \cite{Rubtsova2015,Rubtsova2014}, and rectification \cite{Segal2005,Wu2007,Wu2009}.
More recently, noise \cite{Nicolin2011prb,Nicolin2011,GomezSolano2011,Agarwalla2012}, nonlinear response, e.g., negative differential heat conductance, 
and control by external stimuli \cite{Arrachea2014,Donadio2015} have been examined. An important driving factor in this
growing interest is the development of experimental capabilities that greatly improve on 
the ability to gauge temperatures (and ``effective'' temperatures in nonequilibrium systems) with 
high spatial and thermal resolutions \cite{Huang2006,Tsutsui2008,Hoffman2009,Chen2014,Maher2006,Ioffe2008,Ward2011,Dang2011,Sadat2010,Menges2012,Lee2013,Desiatov2014,Chen2015,Hu2015,Mecklenburg2015}, and to infer from such measurement the 
underlying heat transport processes. In particular, vibrational energy transport/heat conduction in 
molecular layers and junctions has recently been characterized using different probes \cite{Leitner2015,Rubtsova2014,Schwarzer2004,Wang2007,Carter2008,Wang2008,Pein2013,Kasyanenko2011,Meier2014,Kurnosov2015,Yue2015}.

The interplay between charge and energy (electronic and nuclear) transport \cite{Nitzan2007,Galperin2009,Nitzan2011jpcl,Nitzan2011prb,Horsfield2006,DAgosta2008,Asai2011,Asai2015} is of particular interest as it pertains to the performance of energy-conversion devices, 
such as thermoelectric, photovoltaic and electromechanical devices. In particular, the 
thermoelectric response of molecular junctions, mostly focusing on the junction linear response 
as reflected by its Seebeck coefficient, has been recently observed \cite{Reddy2007,Malen2009,Malen2010,Tan2011,Kim2014} and theoretically 
analyzed \cite{Dubi2011,Tan2011,Paulsson2003,Koch2004,Segal2005,Pauly2008,Bergfield2009,Bergfield2010,Ke2009,Liu2009,Ren2012,Wang2011,Lee2014,Amanatidis2015,Simine2015}. Most of the theoretical work has focused on junctions characterized by 
coherent electronic transport in which the electronic and nuclear contribution to heat transport are 
assumed largely independent of each other. The few recent works that analyze electron-phonon 
interactions effects on the junction Seebeck coefficient \cite{Ren2012,Walczak2007,Koch2014,Perroni2014,Zimbovskaya2014} do so in the limit of relatively
weak electron-phonon interaction (in the sense that electron is not localized in the junction), 
using the same level of treatment as applied in the theory of inelastic tunneling spectroscopy. 

The present work considers the opposite limit of strong electron-phonon interaction, 
where electron transport is dominated by successive electron hops subjected to full local 
thermalization, that is, 
successive Marcus electron transfer (ET) processes \cite{Marcus1956,Marcus1964,Marcus1985,Marcus1993,Tachiya1993,Nitzan2006chemical,Peters2015}. 
By their nature, such successive hops are independent of each other, so a single transfer event may be considered. 
Even in this well understood limit different considerations apply under different conditions, and 
different levels of descriptions were applied to account for the molecular nature of the 
solvent \cite{Hynes1989}, 
the dimensionality of the process 
\cite{Tachiya1989,Steeger2015,Schiffer2011,Schiffer2015a,Schiffer2015b,Grunwald1985,Guthrie1996,Lambert2001,Tully2008,Rubtsov2015nature} 
and the definition of the reaction 
coordinate. 
Extensions to equilibrium situations have ranged from considerations of deviation from transition state theory (TST)
to the description of control by external fields \cite{Rubtsov2015nature,Delor2015acr,Delor2015nature}. 

Here, we generalize the standard Marcus (transition state) theory of electron transfer to 
account for situations where the donor and acceptor sites are characterized by different local 
temperatures.  Such generalization requires the use of multidimensional transition state theory 
because nuclear polarization modes associated with the different sites are assumed to be 
equilibrated at their respective local temperatures. Our main results are as follows: (a) We obtain 
an analytical formula for the electron transfer rate that depends on the two site temperatures and 
reduces to the standard Marcus form when these temperatures are equal. (b) The corresponding 
activation energy does not correspond to the geometric activation energy, i.e., the point of lowest 
(free) energy on the isoenergetic surface, and is instead a thermal quantity that depends on the 
the local temperature of each site. (c) Electron
transfer between sites of different temperatures is found to be associated with energy transfer 
between the sites and may affect thermal conduction between sites even when the net electron 
flux between them vanishes. 

We focus on a
model that contains the essential ingredients of our theory: the donor and acceptor sites 
are taken to be at different local temperatures and the electron transfer process is assumed to be 
dominated by two vibrational modes, one localized near the donor and the other near the 
acceptor site at the respective local equilibria. Coupling between these modes that is not 
associated with their mutual coupling to the electron transfer process is disregarded.
The electron transfer rate for this bithermal model is obtained and analyzed, along with 
the implications of this electron transfer process for the energy (heat) transfer between the 
corresponding sites.
While a general treatment of this problem for systems consisting of large numbers of
vibrational modes with associated temperatures 
is tractable, we defer exposition of this formulation to later work. 

\section{\label{sec:Theory}Theory of Electron Transfer between Sites of Different Local Temperatures}
\subsection{Model}
\begin{figure}[t]
\centering
\includegraphics[width = 8.5cm,clip]{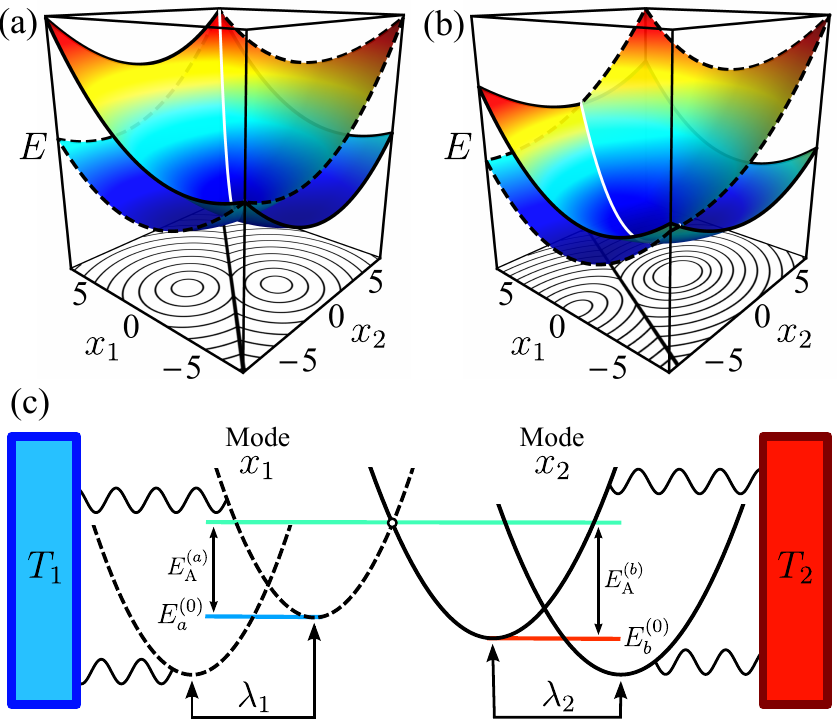}
\caption{\label{fig:Surface}
Energy surfaces ($E_a$ and $E_b$) for electron transfer
between (a) symmetric ($\Delta E_{ba} = 0, E_{\text{R}1} = E_{\text{R}2}$) and (b) asymmetric ($\Delta E_{ba} \neq 0, E_{\text{R}1} \neq E_{\text{R}2}$) donor-acceptor pair geometries.
The boundary of the $E_a$ surface is shown dashed   
and the boundary of the $E_b$ surface is shown as a solid curve.
The $z$-axis corresponds to energy $E$ and is normalized for visual clarity. 
Corresponding contour plots are shown below each surface and 
the crossing line is shown as a thick black line.
(c) Schematic illustration of energy surfaces for electron transfer between modes $x_1$ (dashed) and 
$x_2$ (solid). Each mode is in contact with an independent heat bath. 
The circular marker denotes a crossing  point where $E_a=E_b$.
In this and all other figures, values are shown in dimensionless reduced units.
For convenience, energy may be taken in units of $0.25\,\text{eV}$
(a characteristic reorganization energy) and length in units of $1\,\text{nm}$
(a characteristic donor-acceptor distance).
}
\end{figure}

The system under consideration is similar to the model used in Marcus' theory. It comprises two sites, 1 and 2, on which the transferred electron can localize, and the corresponding electronic states are denoted $a$ (electron on site 1) and $b$ (electron on site 2). The localization is affected by the response of nuclear modes, assumed harmonic, 
whose equilibrium positions depend on the electronic population. 
In the implementation of Marcus' theory, this condition is often expressed in terms of a single reaction coordinate, however the nature of our problem requires the use of at least two groups of modes - one localized near and in (local) thermal equilibrium with site 1, and another localized near and equilibrated with site 2.
In the present discussion 
we consider 
a minimal model comprising two such modes,
denoted $x_1$ and $x_2$, and assume that mode $x_1$ is sensitive to the temperature and charge on site 1 while mode $x_2$ ``feels'' the temperature and charging state of site 2. 
The diabatic electronic (free) energies in states $a$ and $b$ take the same form as in Marcus' theory (see Fig.~\ref{fig:Surface}):
\begin{align}
\label{eq:Ea}
	E_a(x_1,x_2) &= E^{(0)}_a + \frac{1}{2} k_1 (x_1-\lambda_1)^2 + \frac{1}{2} k_2 x_2^2, \\[1ex]
\label{eq:Eb}
	E_b(x_1,x_2) &= E^{(0)}_b + \frac{1}{2} k_1 x_1^2 + \frac{1}{2} k_2 (x_2-\lambda_2)^2.
\end{align}
In choosing these forms we have taken the equilibrium position of mode $x_j: j \in \left\{ 1,2\right\}$ to be at the origin when the corresponding site $j$ is unoccupied. 
A schematic of the geometric and energetic properties for ET using the
considered multidimensional formalism is shown in Fig.~\ref{fig:Surface}(c).
The reorganization energies for each coordinate are
\begin{equation}
	E_{\text{R}1} = \frac{1}{2} k_1 \lambda_1^2 \quad\text{and}\quad E_{\text{R}2} = \frac{1}{2} k_2 \lambda_2^2,
\end{equation}
and the total reorganization energy is 
\begin{equation}
\label{eq:reorg}
	E_\text{R} = E_{\text{R}1}+E_{\text{R}2}.
\end{equation}
As in Marcus theory, we assume that these modes are in thermal equilibrium with their environments, 
however here the environments of sites 1 and 2 are at different local temperatures---$T_1$ and $T_2$---and that modes $x_1$ and $x_2$ are in thermal equilibrium with their corresponding environments.
Our aim is to investigate the effect of this thermal nonequilibrium on the electron transfer process, and to assess the contribution of the latter to the transport of thermal energy between the donor and acceptor sites. In considering the latter, we disregard direct coupling between modes localized near the different sites, so that coupling that may lead to energy transfer between such modes can arise only from their mutual interaction with the electronic subsystem. In reality, heat transport between sites occurs also by direct vibrational coupling.

\subsection{Multidimensional TST}
Because of large disparity between electronic and nuclear timescales, electronic energy conservation is a condition for an electron transfer event to occur. This implies that such events take place only at nuclear configurations that satisfy $E_a(x_1,x_2) = E_b(x_1,x_2)$,
which, denoting $\Delta E_{ba} = E_b^{(0)}-E_a^{(0)}$ and using
Eqs.~(\ref{eq:Ea}) and (\ref{eq:Eb}) can be expressed by the condition $f_\text{c}(x_1,x_2)=0$
where 
\begin{equation}
\label{eq:const}
f_\text{c}(x_1,x_2) = k_1 \lambda_1 x_1 - k_2 \lambda_2 x_2+ \Delta E_{ba} - E_{\text{R}1}+E_{\text{R}2}
\end{equation}
Equation~(\ref{eq:const}) describes a line in the $x_1 \times x_2$ space
on which the two paraboloids displayed in 
Fig.~\ref{fig:Surface}(a) and (b) cross. 
We call this subspace the crossing line (CL).

The Marcus expression for the activation energy is the lowest energy point on this line, and the multidimensional nature of the problem is manifested (in the unithermal case) by an entropic correction to the pre-exponential factor in the rate expression. While this level of description is usually adequate, multidimensional variants of Marcus' theory are developed and applied when a reaction proceeds through complex geometric configurations in which multiple reaction pathways are available \cite{Lambert2001}.
Zwickl \textit{et al.}  \cite{Tully2008} have developed a theory for multiple particle transfer, and have also examined to what extent 
the applicability of a one-dimensional picture persists as the number of intrinsic reaction coordinates is increased.
When a charge transfer reaction occurs through a series of events, 
a univariate parametrization of the reaction progress must often be replaced 
by a set of reaction coordinates to adequately describe the
mechanism \cite{Grunwald1985}. 
For concerted reaction events, 
numerical methods developed by Guthrie have
extended the parabolic Marcus formalism to quartic energy surfaces in hyperdimensional space \cite{Guthrie1996}.
The interplay and competition between sequential and concerted events in ET mechanisms 
has also been investigated, with Lambert \textit{et al.} characterizing 
forbidden and allowed pathways in model systems \cite{Lambert2001}. 
As will be seen below, the fact that different modes affected by the electron transfer represent environments of different temperatures has important implications with regard to the multidimensional nature of the transition state.

\subsection{\label{sec:BTST}Bithermal TST}

Here and below we use the term ``bithermal'' to refer to a two mode model in which the different modes are coupled to environments of different temperatures.
In classical transition state theory for electron transfer that disregards nuclear tunneling the ET rate from state 
$m$ to state $n$ is 
\begin{equation}
\label{eq:genrate}
k_{m \to n} = \tfrac{1}{2}\left\langle\mathcal{T}_{m} v_\perp\right\rangle P_{m \to n}
\end{equation}
where $v_\perp$ is the velocity in the direction normal to the transition surface,
$P_{m \to n}$ is the probability density about the transition state
on the $m$ potential surface calculated at the transition state for the $m \to n$ process,
and $\mathcal{T}_{m}$ is the tunneling probability in the surface crossing event when coming from the $m$ side and is a function of  $v_\perp$ \cite{Schiffer1995,Jonsson2001}.
In the Arrhenius picture,
this expression can be interpreted as a product of
the frequency of reactive attempts multiplied by the probability that an attempt is successful.
Using the Landau-Zener expression for the tunneling probability, we find that
$\mathcal{T}_{m} v_\perp$ is a golden-rule type rate that does not depend on
$v_\perp$
in the weak coupling (nonadiabatic) limit, and is linear in $v_\perp$ in the strong coupling (adiabatic, $\mathcal{T}_{m}=1$) limit (see the Supporting Information). 
For completeness we note that for the two-mode bithermal system considered here, the average velocity in the normal direction is (see the Supporting Information)
\begin{equation}
\left\langle v_\perp \right\rangle = \sqrt{\frac{4}{\pi}\left(\frac{m_2 \beta_2 k_1 E_{\text{R}1} + m_1 \beta_1 k_2 E_{\text{R}2}}{\left| \nabla f_\text{c} \right|^2 m_1 \beta_1 m_2 \beta_2}\right)},
\end{equation}
where $m_j$ is the mass associated with mode $x_j$ and $\left| \nabla f_\text{c} \right|$ is the magnitude of the gradient of the CL constraint.
In the unithermal, equal-mass case ($\beta_1 = \beta_2 = \beta; m_1 = m_2 = m$) this expression reduces to the well-known form
$\sqrt{2/ \pi m \beta}$ 
which is the Boltzmann-weighted expected speed in one-dimension \cite{voth89,Jonsson2001}.
Note however that donor and acceptor sites with significantly different temperatures are far enough from each other to make the nonadiabatic limit the more relevant.

Next consider the probability density $P_{m \to n}$  to be at the transition surface when moving in the $m$ electronic state. In the multidimensional version of Marcus theory this probability is given by the standard activation factor, $\exp\left[-E_\text{A}/k_\text{B} T\right]$ ($k_\text{B}$ is Boltzmann's constant), where the activation energy $E_\text{A}$ is the lowest energy on the transition surface 
multiplied by a pre-exponential term that can be calculated explicitly
(see the Supporting Information). This term will generally also contain entropic corrections that are in in the present harmonic model.
In the multidimensional-bithermal case, the fact that modes of different temperature are weighted differently on the transition surface has to be taken into account. This is accomplished by using Eqs.~(\ref{eq:Ea}) and (\ref{eq:Eb}) to write the required probability density for electronic state $a$ as
\begin{equation}
\begin{aligned}
\label{kabflux}
P_{a \to b}  &=   \iint_{\mathbb{R}^2} \left| \nabla f_\text{c} \right|  e^{-\beta_1 \left(\tfrac{1}{2}k_1 [x_1-\lambda_1]^2\right)} e^{-\beta_2 \left(\tfrac{1}{2}k_2 x_2^2\right)} \\[0ex] 
& \quad \times \delta \big(f_\text{c}(x_1,x_2)\big)  \,dx_1\,dx_2 \\[0ex]
& \quad \Bigg/ \!\! \iint_{\mathbb{R}^2}  e^{-\beta_1 \left(\tfrac{1}{2}k_1 [x_1-\lambda_1]^2\right)} e^{-\beta_2 \left(\tfrac{1}{2}k_2 x_2^2\right)}  \,dx_1\,dx_2\\[1ex]
&= \sqrt{\frac{\beta_1 \beta_2(k_1 E_\text{R1}+ k_2 E_\text{R2})}{2 \pi (\beta_1 E_\text{R2} + \beta_2 E_\text{R1})}}\\[0ex]
& \quad \times \exp{\left[ -\beta_1 \beta_2 \frac{\left(\Delta E_{ba} + E_\text{R}\right)^2}{4 \left(\beta_1 E_\text{R2} + \beta_2 E_\text{R1}\right)} \right]},
\end{aligned}
\end{equation}
and for electronic state $b$,
\begin{equation}
\begin{aligned}
\label{kbaflux}
P_{b \to a} &=   \iint_{\mathbb{R}^2}  \left| \nabla f_\text{c} \right| e^{-\beta_1 \left(\tfrac{1}{2}k_1 x_1^2\right)} e^{-\beta_2 \left(\tfrac{1}{2}k_2 [x_2-\lambda_2]^2\right)} \\[0ex] 
& \quad \times \delta\big(f_\text{c}(x_1,x_2)\big)  \,dx_1\,dx_2 \\[0ex]
& \quad \Bigg/ \!\! \iint_{\mathbb{R}^2}  e^{-\beta_1 \left(\tfrac{1}{2}k_1 x_1^2\right)} e^{-\beta_2 \left(\tfrac{1}{2}k_2 [x_2-\lambda_2]^2\right)}  \,dx_1\,dx_2 \\[1ex]
&= \sqrt{\frac{\beta_1 \beta_2(k_1 E_\text{R1}+ k_2 E_\text{R2})}{2 \pi (\beta_1 E_\text{R2} + \beta_2 E_\text{R1})}} \\[0ex]
& \quad \times \exp{\left[ -\beta_1 \beta_2 \frac{\left(\Delta E_{ba} - E_\text{R}\right)^2}{4 \left(\beta_1 E_\text{R2} + \beta_2 E_\text{R1}\right)} \right]},
\end{aligned}
\end{equation}
where $\beta_j = 1/k_\text{B} T_j$. 
The factor $\left| \nabla f_\text{c} \right|$ renders the constraint $\delta(f_\text{c}(x_1,x_2))$ invariant \cite{Vanden2005,Hartmann2011deltafunction}.
Intervals of integration $\mathbb{R}$ and $\mathbb{R}^2$ 
denote integration over the regions $(-\infty,\infty)$ and $(-\infty,\infty) \times (-\infty,\infty)$,
respectively.

In the relevant nonadiabatic limit, Eqs.~(\ref{kabflux}) and (\ref{kbaflux}) illustrate how the bithermal ET rate 
is related to the inverse thermal energies $\beta_1$ and $\beta_2$ of the respective heat baths. 
Note that they can be written in the standard forms 
\begin{align}
\label{eq:rateexpab}
P_{a \to b} &\propto \exp\left[ -\beta_\text{eff} \frac{\left(\Delta E_{ba} + E_\text{R}\right)^2}{4 E_\text{R} } \right],\\[1ex]
\label{eq:rateexpba}
P_{b \to a} &\propto  \exp\left[ -\beta_\text{eff} \frac{\left(\Delta E_{ba} - E_\text{R}\right)^2}{4 E_\text{R} } \right],
\end{align}
with $\beta_\text{eff} = (k_\text{B} T_\text{eff})^{-1}$, 
where the effective temperature is 
\begin{equation}
\label{eq:Teff}
T_\text{eff} = T_1 \frac{E_\text{R1}}{E_\text{R}}+ T_2\frac{E_\text{R2}}{E_\text{R}}.
\end{equation}
An interesting consequence is that in the symmetric case ($\Delta E_{ba} = 0$) the ratio $P_{a \to b}/ P_{b \to a} = 1$, independent of the site temperatures, so the electron is as likely to reside on either the hot or the cold site. In the unithermal limit ($T_1=T_2 = T$), $T_\text{eff} = T$ and we recover the functional form and temperature dependence predicted by classical Marcus theory \cite{Marcus1956, Thoss2014} (the Supporting Information contains details of this calculation).

Note that one could naively try to evaluate the electron transfer rates by considering the probability to reach the \textit{geometrical} barrier, which is the lowest energy point on the transition surface measured relative to the bottom of the reactant surface. The coordinate of this point can be found by minimizing either $E_a$ or $E_b$ under the constraint $E_a = E_b$. This leads to 
\begin{equation}
\label{eq:x1min}
x^\text{min}_1 = -\lambda_1 \frac{\Delta E_{ba}-E_\text{R}}{2  E_\text{R}} \quad\text{and}\quad x_2^\text{min} = \lambda_2 \frac{\Delta E_{ba}+E_\text{R}}{2  E_\text{R}}.
\end{equation}
The corresponding geometrical activation energies, $E^{(a)}_\text{A}  = E_a(x_1^\text{min},x_2^\text{min})-E^{(0)}_a$ and $E^{(b)}_\text{A}  = E_b(x_1^\text{min},x_2^\text{min})-E^{(0)}_b$  can be cast as additive contributions of energies in mode $x_1$ and in mode $x_2$.  Using Eq.~(\ref{eq:Ea}) we find that for state $a$,
\begin{equation}
E^{(a)}_\text{A} = E^{(a)}_\text{A1}+E^{(a)}_\text{A2} =  \frac{\left(\Delta E_{ba} + E_\text{R}\right)^2}{4 E_\text{R} }, 
\end{equation}
where
\begin{equation}
E^{(a)}_{\text{A}j} = E_{\text{R}j} \left(\frac{\Delta E_{ba}+E_\text{R}}{2  E_\text{R}}\right)^2 : j \in \left\{1,2\right\} 
\end{equation}
Similarly, for state $b$,
\begin{equation}
E^{(b)}_\text{A} = E^{(b)}_\text{A1}+E^{(b)}_\text{A2} =  \frac{\left(\Delta E_{ba} - E_\text{R}\right)^2}{4 E_\text{R} }, 
\end{equation}
and
\begin{equation}
E^{(b)}_{\text{A}j} = E_{\text{R}j} \left(\frac{\Delta E_{ba}-E_\text{R}}{2  E_\text{R}}\right)^2 : j \in \left\{1,2\right\} 
\end{equation}
It follows that the probabilities to reach the configuration $(x_1^\text{min},x_2^\text{min})$ in the $a$ and $b$ states satisfy
\begin{equation}
\label{eq:Pabgeo}
P_{a \to b} \propto \exp \left[ -{\Bigl( \beta_1 E_\text{R1} + \beta_2 E_\text{R2} \Bigl) \left( \frac{\Delta E_{ba} + E_\text{R}}{2 E_\text{R}}\right)^2} \right],
\end{equation}
and
\begin{equation}
\label{eq:Pbageo}
P_{b \to a} \propto  \exp \left[ -{\Bigl( \beta_1 E_\text{R1} + \beta_2 E_\text{R2} \Bigl) \left( \frac{\Delta E_{ba} - E_\text{R}}{2 E_\text{R}}\right)^2} \right],
\end{equation}
which are clearly different from Eqs.~(\ref{kabflux}) and (\ref{kbaflux}), although like the latter they go to the Marcus forms in the limit $\beta_1 = \beta_2$. Interestingly, Eqs.~(\ref{eq:Pabgeo}) and (\ref{eq:Pbageo}) can also be written in the forms (\ref{kabflux}) and (\ref{kbaflux}) but with an effective temperature that satisfies, 
\begin{equation}
\label{eq:Teff2}
\frac{1}{T_\text{eff}} = \frac{1}{T_1} \frac{E_\text{R1}}{E_\text{R}}+ \frac{1}{T_2}\frac{E_\text{R2}}{E_\text{R}},
\end{equation}
an interesting mismatch with Eq.~(\ref{eq:Teff}). 
These differences imply that in the bithermal case the electron transfer rates are no longer controlled by the geometrical barrier. 

\begin{figure}
\centering
\includegraphics[width = 8.5cm,clip]{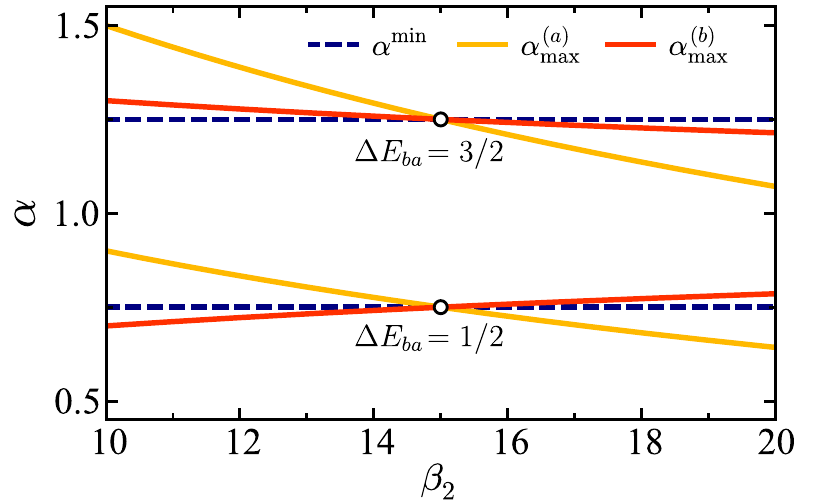}
\caption{\label{fig:Alpha}
Parametric crossing line coordinate $\alpha$ 
shown as function of $\beta_2$, with $\beta_1 = 15$ held constant, for the
geometrical energy minimum (dashed) 
and the maximum probability (solid) on the $E_a$ and $E_b$ surfaces.
In the top curves $\Delta E_{ba} = 3/2$ and in the bottom curves $\Delta E_{ba} = 1/2$.
The circular markers denote the points where $\beta_1 = \beta_2$. 
Other parameters are $E_\text{R1} =  E_\text{R2} = 1/2$.
}
\end{figure}

\begin{figure*}
\centering
\includegraphics[width = 17.0cm,clip]{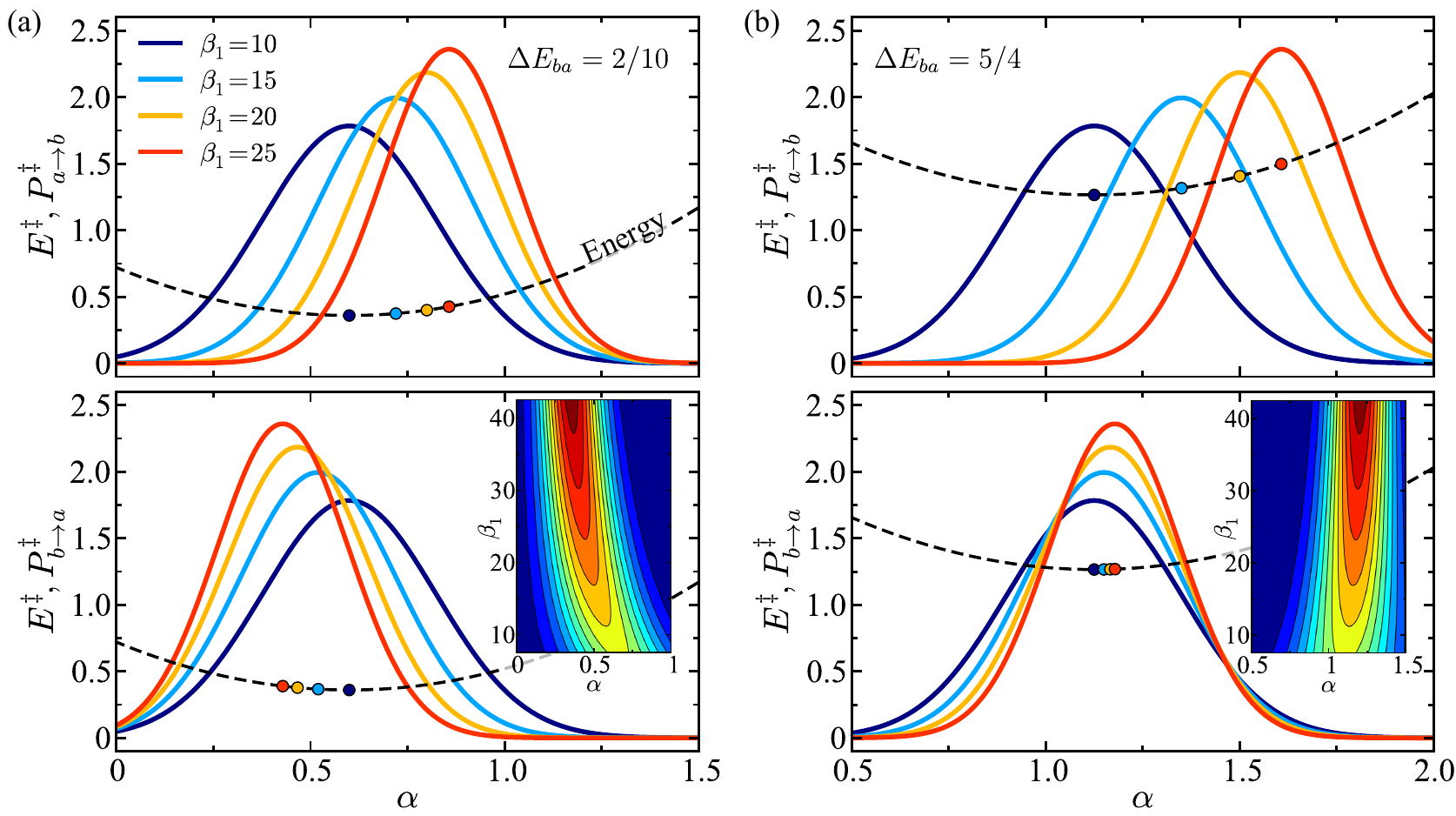}
\caption{\label{fig:EnergyProb}
Crossing point probability densities $P^\ddag (\alpha)$ 
for (a) $\Delta E_{ba}=2/10$ and (b) $\Delta E_{ba}=5/4$ on the $E_a$ (top) and $E_b$ (bottom) energy surfaces
as functions of the crossing line coordinate $\alpha$ (Eq.~(\ref{eq:para})).
Varying values of $\beta_1$ are shown with $\beta_2 = 10$ held constant in all cases.
In each panel, the corresponding crossing line energy $E^\ddag$ is shown as a parabolic dashed curve.
The circular markers on the energy curves 
denote the corresponding thermal energy minima (probability density maxima).
In each bottom panel, the inset is a corresponding contour plot of $P_{b \to a}^\ddag (\alpha)$ 
which is normalized with colors varying from blue (min) to red (max).
Other parameters are $E_\text{R1} = E_\text{R2}=1/2$.
}
\end{figure*}

This can be also seen explicitly: The equal electronic energies condition defines the CL, which can be parametrized in terms of a coordinate $\alpha$ according to
\begin{equation}
\begin{aligned}
\label{eq:para}
x_1(\alpha) &= \frac{k_2 \lambda_2}{k_1 \lambda_1} \alpha + \frac{1}{k_1 \lambda_1}\left( \frac{1}{2} k_1 \lambda_1^2 - \frac{1}{2} k_2 \lambda_2^2 - \Delta E_{ba}\right), \\[1ex]
x_2(\alpha) &= \alpha.
\end{aligned}
\end{equation}
with a value of the parametric coordinate $\alpha$ specifying a unique transition point.
The energy on the CL,
\begin{equation} 
E^\ddag(\alpha) = E_a[x_1(\alpha),x_2(\alpha)] = E_b[x_1(\alpha),x_2(\alpha)],
\end{equation} is parametrized by $\alpha$.
The energies as a function of position $\alpha$ on the crossing line coming from states $a$ and $b$, relative to the corresponding energy origins are
\begin{align}
E^\ddag(\alpha) - E_a^{(0)} &= \frac{1}{2}k_1 [x_1(\alpha)-\lambda_1]^2+\frac{1}{2}k_2 [x_2(\alpha)]^2, \\[1ex]
E^\ddag(\alpha) - E_b^{(0)} &= \frac{1}{2}k_1 [x_1(\alpha)]^2+\frac{1}{2}k_2 [x_2(\alpha)-\lambda_2]^2,
\end{align} 
respectively.
The probabilities to be at point $\alpha$ on the CL 
given that we are in the corresponding state satisfy
\begin{align}
\label{eq:probdensab}
P^\ddag_{a \to b}(\alpha) &= \frac{\displaystyle e^{{-\beta_1 \left(\tfrac{1}{2}k_1 [x_1(\alpha)-\lambda_1]^2\right)}} e^{{-\beta_2 \left(\tfrac{1}{2}k_2 [x_2(\alpha)]^2\right)}}}{\displaystyle \int_{\mathbb{R}}  e^{{-\beta_1 \left(\tfrac{1}{2}k_1 [x_1(\alpha)-\lambda_1]^2\right)}} e^{{-\beta_2 \left(\tfrac{1}{2}k_2 [x_2(\alpha)]^2\right)}} \,d\alpha}, \\[1ex]
\label{eq:probdensba}
P^\ddag_{b \to a}(\alpha) &= \frac{\displaystyle e^{{-\beta_1 \left(\tfrac{1}{2}k_1 [x_1(\alpha)]^2\right)}} e^{{-\beta_2 \left(\tfrac{1}{2}k_2 [x_2(\alpha)-\lambda_2]^2\right)}}}{\displaystyle \int_{\mathbb{R}}  e^{{-\beta_1 \left(\tfrac{1}{2}k_1 [x_1(\alpha)]^2\right)}} e^{{-\beta_2 \left(\tfrac{1}{2}k_2 [x_2(\alpha)-\lambda_2]^2\right)}} \,d\alpha}.
\end{align}
For $P^\ddag_{a \to b}(\alpha)$, the point of maximum probability on the CL is found from Eq.~(\ref{eq:probdensab}) to be 
\begin{align}
\label{eq:x1maxa}
x_{1,\text{max}}^{(a)} &=  \frac{\lambda_1\left[\beta_2(-\Delta E_{ba}+E_\text{R1})-(\beta_2-2\beta_1)E_\text{R2}\right]}{2(E_\text{R2}\beta_1 + E_\text{R1}\beta_2)},\\[1ex]
\label{eq:x2maxa}
x_{2,\text{max}}^{(a)} &= \alpha^{(a)}_{\text{max}}= \frac{\lambda_2 \beta_1 (\Delta E_{ba}+E_\text{R})}{2(E_\text{R2}\beta_1 + E_\text{R1}\beta_2)}. 
\end{align}
A similar procedure using Eq.~(\ref{eq:probdensba}) yields
\begin{align}
\label{eq:x1maxb}
x_{1,\text{max}}^{(b)} &= \frac{\lambda_1 \beta_2 (-\Delta E_{ba}+E_\text{R})}{2(E_\text{R2}\beta_1 + E_\text{R1}\beta_2)},  \\[1ex]
\label{eq:x2maxb}
x_{2,\text{max}}^{(b)} &= \alpha^{(b)}_{\text{max}} =  \frac{\lambda_2\left[\beta_1(\Delta E_{ba}+E_\text{R2})-(\beta_1-2\beta_2)E_\text{R1}\right]}{2(E_\text{R2}\beta_1 + E_\text{R1}\beta_2)}.
\end{align}
For $\beta_1 = \beta_2$, the position of maximum probability is also the geometric minimum. When the temperatures differ, the position of maximum probability on the transition line shifts from this minimum. The shifts of these probability distributions from their unithermal forms is the reason for the difference between the correct probabilities given by Eqs.~(\ref{kabflux}) and (\ref{kbaflux}), and the forms in Eqs.~(\ref{eq:Pabgeo}) and (\ref{eq:Pbageo}) obtained under the assumption that the probabilities are dominated by the geometric minimum energy.
A graphical representation of these results is shown in Figs.~\ref{fig:Alpha}
and \ref{fig:EnergyProb} for several illustrative examples. Figure~\ref{fig:Alpha} shows the position of maximum probability as a function of the temperature difference. The probability densities themselves are shown in Fig.~\ref{fig:EnergyProb}. These plots clearly show the essentials of the bithermal transition behavior as discussed above.

\begin{figure}
\centering
\includegraphics[width = 8.5cm,clip]{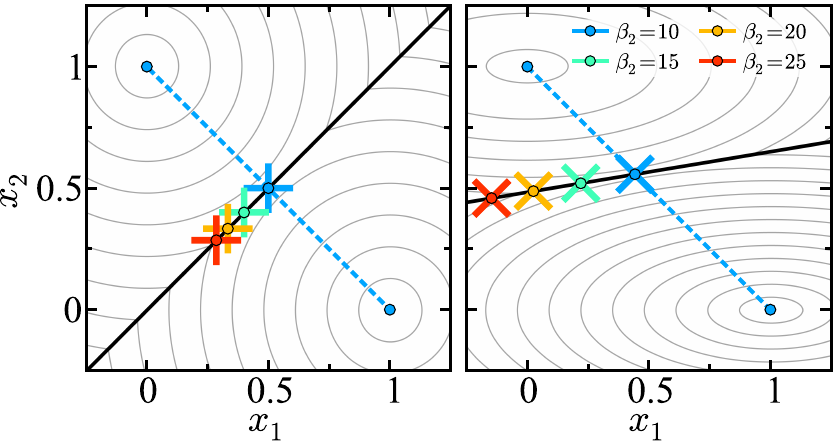}
\caption{\label{fig:MinContour}
Contour plots of energy surfaces for
symmetric (left) and asymmetric (right) donor-acceptor pair geometries.
The crossing line is shown as a thick black line.
The crosses mark the point of maximum probability for the $a \to b$ transition on the crossing line for
$\beta_2 \in \left\{10,15,20,25\right\}$ with $\beta_1 = 10$ held constant.
The dashed line connects the two well minima 
through the geometrical minimum energy point.
}
\end{figure}

The following observations are noteworthy:
\begin{enumerate}[(a)] 
\item The point of maximum probability on the transition surface does not depend on the absolute temperatures $T_1$ and $T_2$, only on their ratios. When $T_1=T_2$ it becomes the geometrical point of minimum enegy which is temperature independent. 
\item Considering the position of the maximum probability points relative to the minimum energy point on the CL, some general trends can observed. For reaction free energies below the total reorganization energy ($|E_{ba}|<E_\text{R}$) the points of maximum probability in the $a \to b$ and $b \to a$ directions are on opposite sides of the geometrical energy minimum for $\beta_2<\beta_1$, cross at the unithermal point, and finally continue on opposite sides for $\beta_2>\beta_1$. For reactions with reorganization energy above the reaction free energy ($|E_{ba}|>E_\text{R}$) the maximum probability points for both reaction directions are on same side of the geometrical energy minimum for all values of $\beta_2$ with $\beta_1$ held constant, except where they cross at the unithermal point. 
\item As shown in Fig.~\ref{fig:EnergyProb}, in addition to the shift in the transition line probability distribution function, another interesting feature is observed; both the $P^\ddag_{a \to b}$  and $P^\ddag_{b \to a}$ distributions become narrower (smaller variance) with increasing deviation from the unithermal point in the direction $\beta_1>\beta_2$ for finite $\beta_2$ held constant.
The inset in each bottom panel of Fig.~\ref{fig:EnergyProb} illustrates this narrowing as $\beta_1 \to \infty$.
In the opposite direction ($\beta_1<\beta_2$), the complementary trend is observed with the distributions becoming increasingly broad.
It is of note that in the limit $\beta_1 \to 0$ $(T_1 \to \infty)$ the total distribution will be dominated by the respective distribution of the $x_2$ coordinate, i.e.,
$P^\ddag_{a \to b}(x_1,x_2) \approx P^\ddag_{a \to b}(x_2)$.
\item At the unithermal limit, the maximum probability path that connects stable states is linear and goes through $\alpha^\text{min}$  as shown in Fig.~\ref{fig:EnergyProb}. This holds in both the symmetric ($E_{ba} = 0$) and asymmetric cases. In bithermal systems, this path is obviously nonlinear (since it deviates from the minimum energy point) and depends on the thermal characteristics. Fig.~\ref{fig:MinContour} demonstrates this observation.
Note that unlike in the symmetric case, in an asymmetric system the path connecting minima is not necessarily normal to the CL.
This is also the case in unithermal charge transfer reactions with asymmetric donor-acceptor geometry \cite{Newton2015}.
The finding of a thermal energy minimum point 
that does not correspond to a geometrical energy minimum point 
is nonintuitive, but is congruent with recent advances 
in transition state theory which have shown
that in nonequilibrium systems
the traditional picture of a transition state 
as a stationary saddle point on a potential energy surface 
is flawed, and that the correct nature 
is a structure with different extremal properties \cite{dawn05a,craven14a,craven14c,craven15a,craven15c}. 
\end{enumerate}

Finally, an interesting interpretation of the results (\ref{kabflux}) and (\ref{kbaflux}) can be found in terms of the Tolman activation energy \cite{Tolman1920} that accounts for statistical properties of the reaction mechanism and goes beyond the Arrhenius viewpoint of a single activation threshold. In the Tolman interpretation, the activation energy is defined as the average energy of all reacting systems minus the average energy of all reactants  \cite{Tolman1920,Truhlar1978,Truhlar2001}. 
In the present model this is
\begin{equation}
E^{\text{Tolman},(m)}_\text{A} = \Big\langle E^\ddag (\alpha) \Big\rangle_m- E^{(0)}_m : m \in \left\{a,b\right\}
\end{equation}
where $E^\ddag (\alpha)$ is the energy on the CL and the average is over the corresponding distribution ($m \in \left\{a,b\right\}$), namely
\begin{equation}
   \Big\langle E^\ddag (\alpha) \Big\rangle_m = \int_{\mathbb{R}}  E^\ddag (\alpha)  \, P^\ddag_{m \to n}(\alpha) \,d\alpha. \label{eq:actenergyint} 
\end{equation}
Using Eqs.~(\ref{eq:probdensab}) and (\ref{eq:probdensba})
these averages can be easily evaluated and can be cast as additive terms representing the division of the needed activation energy between modes $x_1$ and $x_2$,
\begin{equation}
E^{\text{Tolman},(m)}_\text{A}=  \Big\langle E^{(m)}_\text{A1} \Big\rangle + \Big\langle E^{(m)}_\text{A2} \Big\rangle : m \in \left\{a,b\right\},
\end{equation}
where
\begin{equation}
\begin{aligned}
 \Big\langle E^{(a)}_\text{A1} \Big\rangle &= \frac{2 \beta_1 E_\text{R2}^2 + 2 \beta_2 E_\text{R1}E_\text{R2}+ \beta_2^2 E_\text{R1} (\Delta E_{ba}+E_\text{R})^2}{4(E_\text{R2}\beta_1 + E_\text{R1}\beta_2)^2}, \\[1ex]
 \Big\langle E^{(a)}_\text{A2} \Big\rangle &= \frac{2 \beta_2 E_\text{R1}^2 + 2 \beta_1 E_\text{R1}E_\text{R2}+ \beta_1^2 E_\text{R2} (\Delta E_{ba}+E_\text{R})^2}{4(E_\text{R2}\beta_1 + E_\text{R1}\beta_2)^2}, \\[1ex]
 \Big\langle E^{(b)}_\text{A1} \Big\rangle &= \frac{2 \beta_1 E_\text{R2}^2 + 2 \beta_2 E_\text{R1}E_\text{R2}+ \beta_2^2 E_\text{R1} (\Delta E_{ba}-E_\text{R})^2}{4(E_\text{R2}\beta_1 + E_\text{R1}\beta_2)^2},\\[1ex]
 \Big\langle E^{(b)}_\text{A2} \Big\rangle &= \frac{2 \beta_2 E_\text{R1}^2 + 2 \beta_1 E_\text{R1}E_\text{R2}+ \beta_1^2 E_\text{R2} (\Delta E_{ba}-E_\text{R})^2}{4(E_\text{R2}\beta_1 + E_\text{R1}\beta_2)^2}. \\[1ex]
\end{aligned}
\end{equation}
It can be easily checked that defining the probabilities to be on the CL by  
\begin{equation} 
P_{a \to b} \propto \exp \left[ -\Bigl( \beta_1 \Big\langle E^{(a)}_\text{A1} \Big\rangle + \beta_2 \Big\langle E^{(a)}_\text{A2} \Big\rangle \Bigl) \right],
\end{equation}
and
\begin{equation} 
P_{b \to a} \propto \exp \left[ -\Bigl( \beta_1 \Big\langle E^{(b)}_\text{A1} \Big\rangle + \beta_2 \Big\langle E^{(b)}_\text{A2} \Big\rangle \Bigl) \right],
\end{equation}
leads to the exact results (\ref{kabflux}) and (\ref{kbaflux}) for the bithermal Boltzmann factors.

\section{\label{sec:ET}Energy Transfer}
As outlined in the introduction, the coupled transfer of charge and heat, and the interplay between the 
electric and heat currents, gives rise to unique electronic and thermoelectric phenomena \cite{Esposito2015, Lim2013}.
When electron transfer takes place across a thermal gradient,
it can carry energy as well, implying heat ($\mathcal{Q}$) transfer between the donor and acceptor sites. Indeed, our model has disregarded direct coupling between the modes coupled to the electronic occupation of the different sites, so this coupling is the only potential source (in this model) of heat transfer. Here we explore this possibility.

During the $m \to n$ state transition,  for mode $x_j$, 
the heat transferred is the sum of the heat released by the corresponding bath
during the ascent to the transition state crossing point defined by $\alpha$ on the $E_m$ surface,
and the heat absorbed by the bath 
during the descent to equilibrium on the $E_n$ surface, 
\begin{equation}
\mathcal{Q}^{(m \to n)}_j(\alpha) = - \mathcal{Q}^{(m)}_\text{rel} + \mathcal{Q}^{(n)} _\text{abs}.
\end{equation}
For the two-mode two-state system considered here
the amounts of heat transfer into each bath during an electron transfer event are: 
\begin{equation}
\begin{aligned}
	\label{eq:atob}
   \mathcal{Q}^{(a\to b)}_1(\alpha)  &=-\mathcal{Q}^{(b \to a)}_1(\alpha)  \\[1ex]
	&=  -\tfrac{1}{2} k_1 [x_1(\alpha)-\lambda_1]^2 + \tfrac{1}{2} k_1 [x_1(\alpha)]^2, \\[1ex]
	\mathcal{Q}^{(a \to b)}_2(\alpha)  &=-\mathcal{Q}^{(b \to a)}_2(\alpha)  \\[1ex]
	&=  -\tfrac{1}{2} k_2 [x_2(\alpha)]^2 + \tfrac{1}{2} k_2 [x_2(\alpha)-\lambda_2]^2. \\[1ex]
\end{aligned}
\end{equation}
The signs in Eq.~(\ref{eq:atob})
are chosen such that
$\mathcal{Q}$ is positive when energy enters the corresponding bath.
The average values for these components are
\begin{equation}
\begin{aligned}
  \Big\langle \mathcal{Q}^{(a \to b)}_j \Big\rangle &= \int_{\mathbb{R}}    \mathcal{Q}^{(a \to b)}_j(\alpha) \, P^\ddag_{a \to b}(\alpha) \,d\alpha,\\[1ex]
	\Big\langle \mathcal{Q}^{(b \to a)}_j \Big\rangle &= \int_{\mathbb{R}}    \mathcal{Q}^{(b \to a)}_j(\alpha)  \,P^\ddag_{b \to a}(\alpha) \,d\alpha,
\end{aligned}
\end{equation}
where $j \in \left\{1,2\right\}$ 
and  $P^\ddag_{m \to n}(\alpha)$ is the probability density on the CL for the corresponding surface.
Evaluating each of these integrals yields
\begin{equation}
\begin{aligned}
\label{eq:heatcomp}
 \Big\langle \mathcal{Q}^{(a \to b)}_1 \Big\rangle &= \frac{-E_\text{R1} T_1 \Delta E_{ba} + E_\text{R1} E_\text{R2}(T_2-T_1)}{E_\text{R1} T_1 + E_\text{R2}T_2},\\[1ex]
 \Big\langle \mathcal{Q}^{(a \to b)}_2 \Big\rangle &= \frac{-E_\text{R2} T_2 \Delta E_{ba} - E_\text{R1}  E_\text{R2}(T_2-T_1)}{E_\text{R1} T_1 + E_\text{R2}T_2},\\[1ex]
 \Big\langle \mathcal{Q}^{(b \to a)}_1 \Big\rangle &= \frac{E_\text{R1} T_1 \Delta E_{ba} + E_\text{R1} E_\text{R2}(T_2-T_1)}{E_\text{R1} T_1 + E_\text{R2}T_2},\\[1ex]
 \Big\langle \mathcal{Q}^{(b \to a)}_2 \Big\rangle &= \frac{E_\text{R2} T_2 \Delta E_{ba} - E_\text{R1} E_\text{R2}(T_2-T_1)}{E_\text{R1} T_1 + E_\text{R2}T_2},
\end{aligned}
\end{equation}
which depend on 
the reaction free energy,
the reorganization energy in each mode, 
and the temperature of each bath. 
It should be emphasized that the modes themselves are assumed to remain in thermal
equilibrium. 
Expressions (\ref{eq:heatcomp}) give the heat transferred into the thermal bath with which the
corresponding mode equilibrates for a single electron transfer in the indicated direction. Note that
the total heat transfer for the $a \to b$ transition is
\begin{equation}
 \Big\langle \mathcal{Q}^{(a \to b)} \Big\rangle = \Big\langle \mathcal{Q}^{(a \to b)}_1 \Big\rangle  + \Big\langle \mathcal{Q}^{(a \to b)}_2 \Big\rangle 
																				 	 = -\Delta E_{ba},
\end{equation}
and correspondingly for the $b \to a$ transition,
\begin{equation}
 \Big\langle \mathcal{Q}^{(b \to a)} \Big\rangle = \Big\langle \mathcal{Q}^{(b \to a)}_1 \Big\rangle  + \Big\langle \mathcal{Q}^{(b \to a)}_2 \Big\rangle 
																				 	 = \Delta E_{ba},
\end{equation}
which are just statements of energy conservation.
The change in free energy of the baths associated with the $a \to b$ process ($-\Delta E_{ba}$)is divided between the two baths 
with the ratio $E_\text{R1} T_1 / E_\text{R2} T_2$.
Interestingly, this ratio depends on their temperatures, reflecting the
fact that the higher temperature bath is more effective in promoting electron transfer. 
Even more significant is the observation that there is a term in each expression in (\ref{eq:heatcomp}) that does not depend on
$\Delta E_{ba}$, and the sign of which does not depend on the direction of the electron transfer process. Thus, there exists a nonzero heat transfer
between baths associated with the electron transfer process in bithermal systems. Over each electron transfer event it is given by
\begin{equation}
\label{eq:heattot1}
\Big\langle \mathcal{Q}_{2 \to 1} \Big\rangle \equiv \Big\langle \mathcal{Q}^{(a \to b)}_1 \Big\rangle  + \Big\langle \mathcal{Q}^{(b \to a)}_1 \Big\rangle = \frac{2 E_{R1} E_\text{R2} (T_2-T_1)}{E_\text{R1}T_1 + E_\text{R2}T_2},
\end{equation}
and 
\begin{equation}
\label{eq:heattot2}
\Big\langle \mathcal{Q}_{1 \to 2} \Big\rangle \equiv \Big\langle \mathcal{Q}^{(a \to b)}_2 \Big\rangle  + \Big\langle \mathcal{Q}^{(b \to a)}_2 \Big\rangle = -\frac{2 E_{R1} E_\text{R2} (T_2-T_1)}{E_\text{R1}T_1 + E_\text{R2}T_2}.
\end{equation}

To see the significance of this result, consider an ensemble of
site-pairs with probabilities $p_a$ that a pair is in state $a$ (electron on site 1)
and $p_b$ that the pair is in state $b$ (electron on site 2). These probabilities obey the kinetic equations
\begin{equation}
\frac{d p_a}{d t} = -\frac{d p_b}{d t} = - \mathcal{J}_{a \to b}+
\mathcal{J}_{b \to a}, 
\end{equation}
where $\mathcal{J}_{a \to b} = k_{a \to b} p_a$ and 
$\mathcal{J}_{b \to a} = k_{b \to a} p_b$.
Correspondingly, the rate of heat deposit on the respective site is given by
\begin{equation}
\frac{d \mathcal{Q}_j}{d t} = \mathcal{J}_{a \to b} \Big\langle \mathcal{Q}_j^{(a \to b)} \Big\rangle + \mathcal{J}_{b \to a} \Big\langle \mathcal{Q}_j^{(b \to a)} \Big\rangle : j \in \left\{1,2\right\}.
\end{equation}
Now consider the steady state at which the system is at electronic quasiequilibrium so that $\mathcal{J}_{a \to b} = \mathcal{J}_{b \to a} = \mathcal{J}_\text{ss}$, i.e., the net electron flux between sites vanishes. Using Eqs.~(\ref{eq:heattot1}) and (\ref{eq:heattot2})
it follows that at this state
\begin{equation}
\label{eq:heatflux}
\bigg(\frac{d \mathcal{Q}_1}{d t}\bigg)_{\!\text{ss}} = -\bigg(\frac{d \mathcal{Q}_2}{d t}\bigg)_{\!\text{ss}} = \mathcal{J}_\text{ss}\frac{2 E_{R1} E_\text{R2} (T_2-T_1)}{E_\text{R1}T_1 + E_\text{R2}T_2} \equiv \mathcal{J}_\text{ss}^\mathcal{Q}.
\end{equation}
Thus, for $T_1 \neq T_2$,
even when the net electron flux vanishes, 
the presence of hopping electrons induces
a net heat current from the hot bath to the cold bath.
Of interest is the observation 
that there is no pure Seebeck effect in the model investigated here.
This is seen in Eqs.~(\ref{eq:rateexpab})-(\ref{eq:Teff}) which imply that when $E_{\text{R}1}=E_{\text{R}2}$, 
changing $T_1$ relative to $T_2$ affects the forward and backward rates in the same way.
Note that Eq.~(\ref{eq:heatflux})
is nonlinear in the temperature difference (although it is approximately so when the difference is small).
In the  high- and low-temperature limits of 
site 2, the steady-state heat flux becomes
\begin{equation}
\lim_{T_2\to \infty} \mathcal{J}^\mathcal{Q}_\text{ss}   = 2\mathcal{J}_\text{ss} E_{R1} \quad\text{and}\quad \lim_{T_2\to 0}  \mathcal{J}^{\mathcal{Q}}_\text{ss}  = -2 \mathcal{J}_\text{ss} E_{R2}, 
\end{equation}
respectively,
which each depend only on the reorganization energy of the respective cold mode.
These results imply that in a system where electron hops between local sites, there is a contribution to the heat conduction associated with the electronic motion. An assessment of this contribution to the heat conduction in such systems will be made elsewhere.

\section{\label{sec:Conc}Conclusions}
A unified theory for the rate and extent of electron  transfer
and heat transport between bithermal donor-acceptor pairs 
has been constructed in an augmented Marcus framework.
Through application of a multidimensional 
transition state theory where different modes interact with environments of different temperatures, 
we have characterized the kinetics of the charge transfer process 
over various temperature gradients and  geometries between reactant and product states.
In a bithermal system,
the traditional interpretation of the activation energy as a  single-point 
derived through geometric minimization of over all points where 
the donor and acceptor are equienergetic has been shown to not adequately describe the transfer mechanism, and
instead,
a statistical interpretation of the activation energy threshold has been developed 
to account for the biasing of states that arises due to the temperature gradient.
We find that entropic rate corrections, which are trivial in the unithermal case, 
are nontrivial for bithermal systems
and are characteristic of the multithermal density of states.
Surprisingly, for electron transport across a thermal gradient, 
the transfer of heat continues to occur even when there is no net transfer of charge.
This effect could be harnessed, particularly through molecular junctions and wires \cite{Nitzan2007jpcm,Nitzan2007,Nitzan2010,Chet2013},
to control the transfer of thermal energy in reaction networks with complex systems of heat reservoirs.
In turn, the use of these reservoirs to control charge current in thermoelectric systems with
nonzero Seebeck coefficients could result in the development of devices and electronics that can be harnessed for 
application in thermally controlled molecular machines.

A description of the transfer process across smoothly-varying temperature gradients,
and the characterization of possible deviations from the assumed bithermal Boltzmann distribution
on the transition state crossing line are possible areas for future research.
The treatment of collective behaviors arising from anharmonic coupling between reactive modes, 
such as that observed in multiple particle transfer mechanisms \cite{Tully2008}, 
will require further characterization of the nature of thermalization \cite{hern07a} and temperature, specifically 
in systems that are in contact with multiple independent heat baths.
The current description gives impetus for experimental verification of the 
constructed methodologies in bithermal systems.

The bithermal donor-acceptor model considered here 
can be generalized to systems with multiple reaction pathways. 
For example, a theoretical description of the transfer mechanism in a donor-bridge-acceptor model
can be constructed by extending the dimension of the transition state structure on the crossing ``line''.
Developing a general description of thermal transition states in 
electron transfer reactions with many reactive modes
could be accomplished through implementation of 
the geometric transition state formalisms developed for classical reactions in high dimensionality \cite{Uzer02}. 
A conjecture supported by the bithermal biasing of the transition state structure predicted here 
is that multi-body temperature gradients can be used to control which reaction pathway is taken in a complex network. 
The possibility of controlling reactions through multithermally-induced deformation of transitions states 
is a significant finding of this study, and one that is primed for further exploration thorough computation and experiment.

\section{\label{sec:Ack}Acknowledgements}
AN's research is supported by the Israel Science Foundation, the US-Israel Bi-national Science Foundation and the University of Pennsylvania.

\bibliography{c1.bbl}

\begin{thebibliography}{126}
\expandafter\ifx\csname natexlab\endcsname\relax\def\natexlab#1{#1}\fi
\expandafter\ifx\csname bibnamefont\endcsname\relax
  \def\bibnamefont#1{#1}\fi
\expandafter\ifx\csname bibfnamefont\endcsname\relax
  \def\bibfnamefont#1{#1}\fi
\expandafter\ifx\csname citenamefont\endcsname\relax
  \def\citenamefont#1{#1}\fi
\expandafter\ifx\csname url\endcsname\relax
  \def\url#1{\texttt{#1}}\fi
\expandafter\ifx\csname urlprefix\endcsname\relax\def\urlprefix{URL }\fi
\providecommand{\bibinfo}[2]{#2}
\providecommand{\eprint}[2][]{\url{#2}}

\bibitem[{\citenamefont{Galperin
  et~al.}(2007{\natexlab{a}})\citenamefont{Galperin, Ratner, and
  Nitzan}}]{Nitzan2007jpcm}
\bibinfo{author}{\bibfnamefont{M.}~\bibnamefont{Galperin}},
  \bibinfo{author}{\bibfnamefont{M.~A.} \bibnamefont{Ratner}},
  \bibnamefont{and} \bibinfo{author}{\bibfnamefont{A.}~\bibnamefont{Nitzan}},
  \bibinfo{journal}{J. Phys.: Condens. Matter} \textbf{\bibinfo{volume}{19}},
  \bibinfo{pages}{103201} (\bibinfo{year}{2007}{\natexlab{a}}).

\bibitem[{\citenamefont{Dubi and Di~Ventra}(2011)}]{Dubi2011}
\bibinfo{author}{\bibfnamefont{Y.}~\bibnamefont{Dubi}} \bibnamefont{and}
  \bibinfo{author}{\bibfnamefont{M.}~\bibnamefont{Di~Ventra}},
  \bibinfo{journal}{Rev. Mod. Phys.} \textbf{\bibinfo{volume}{83}},
  \bibinfo{pages}{131} (\bibinfo{year}{2011}).

\bibitem[{\citenamefont{Cahill et~al.}(2002)\citenamefont{Cahill, Goodson, and
  Majumdar}}]{Cahill2002}
\bibinfo{author}{\bibfnamefont{D.~G.} \bibnamefont{Cahill}},
  \bibinfo{author}{\bibfnamefont{K.}~\bibnamefont{Goodson}}, \bibnamefont{and}
  \bibinfo{author}{\bibfnamefont{A.}~\bibnamefont{Majumdar}},
  \bibinfo{journal}{J. Heat Transfer} \textbf{\bibinfo{volume}{124}},
  \bibinfo{pages}{223} (\bibinfo{year}{2002}).

\bibitem[{\citenamefont{Cahill et~al.}(2003)\citenamefont{Cahill, Ford,
  Goodson, Mahan, Majumdar, Maris, Merlin, and Phillpot}}]{Cahill2003}
\bibinfo{author}{\bibfnamefont{D.~G.} \bibnamefont{Cahill}},
  \bibinfo{author}{\bibfnamefont{W.~K.} \bibnamefont{Ford}},
  \bibinfo{author}{\bibfnamefont{K.~E.} \bibnamefont{Goodson}},
  \bibinfo{author}{\bibfnamefont{G.~D.} \bibnamefont{Mahan}},
  \bibinfo{author}{\bibfnamefont{A.}~\bibnamefont{Majumdar}},
  \bibinfo{author}{\bibfnamefont{H.~J.} \bibnamefont{Maris}},
  \bibinfo{author}{\bibfnamefont{R.}~\bibnamefont{Merlin}}, \bibnamefont{and}
  \bibinfo{author}{\bibfnamefont{S.~R.} \bibnamefont{Phillpot}},
  \bibinfo{journal}{J. Appl. Phys.} \textbf{\bibinfo{volume}{93}},
  \bibinfo{pages}{793} (\bibinfo{year}{2003}).

\bibitem[{\citenamefont{Leitner}(2008)}]{Leitner2008}
\bibinfo{author}{\bibfnamefont{D.~M.} \bibnamefont{Leitner}},
  \bibinfo{journal}{Annu. Rev. Phys. Chem.} \textbf{\bibinfo{volume}{59}},
  \bibinfo{pages}{233} (\bibinfo{year}{2008}).

\bibitem[{\citenamefont{Leitner}(2015)}]{Leitner2015}
\bibinfo{author}{\bibfnamefont{D.~M.} \bibnamefont{Leitner}},
  \bibinfo{journal}{Adv. Phys.} \textbf{\bibinfo{volume}{64}},
  \bibinfo{pages}{445} (\bibinfo{year}{2015}).

\bibitem[{\citenamefont{Li et~al.}(2012)\citenamefont{Li, Ren, Wang, Zhang,
  H\"anggi, and Li}}]{Li2012}
\bibinfo{author}{\bibfnamefont{N.}~\bibnamefont{Li}},
  \bibinfo{author}{\bibfnamefont{J.}~\bibnamefont{Ren}},
  \bibinfo{author}{\bibfnamefont{L.}~\bibnamefont{Wang}},
  \bibinfo{author}{\bibfnamefont{G.}~\bibnamefont{Zhang}},
  \bibinfo{author}{\bibfnamefont{P.}~\bibnamefont{H\"anggi}}, \bibnamefont{and}
  \bibinfo{author}{\bibfnamefont{B.}~\bibnamefont{Li}}, \bibinfo{journal}{Rev.
  Mod. Phys.} \textbf{\bibinfo{volume}{84}}, \bibinfo{pages}{1045}
  (\bibinfo{year}{2012}).

\bibitem[{\citenamefont{Dhar}(2008)}]{Dhar2008}
\bibinfo{author}{\bibfnamefont{A.}~\bibnamefont{Dhar}}, \bibinfo{journal}{Adv.
  Phys.} \textbf{\bibinfo{volume}{57}}, \bibinfo{pages}{457}
  (\bibinfo{year}{2008}).

\bibitem[{\citenamefont{Luo and Chen}(2013)}]{Luo2013}
\bibinfo{author}{\bibfnamefont{T.}~\bibnamefont{Luo}} \bibnamefont{and}
  \bibinfo{author}{\bibfnamefont{G.}~\bibnamefont{Chen}},
  \bibinfo{journal}{Phys. Chem. Chem. Phys.} \textbf{\bibinfo{volume}{15}},
  \bibinfo{pages}{3389} (\bibinfo{year}{2013}).

\bibitem[{\citenamefont{Rubtsova
  et~al.}(2015{\natexlab{a}})\citenamefont{Rubtsova, Qasim, Kurnosov, Burin,
  and Rubtsov}}]{Rubtsova2015acr}
\bibinfo{author}{\bibfnamefont{N.~I.} \bibnamefont{Rubtsova}},
  \bibinfo{author}{\bibfnamefont{L.~N.} \bibnamefont{Qasim}},
  \bibinfo{author}{\bibfnamefont{A.~A.} \bibnamefont{Kurnosov}},
  \bibinfo{author}{\bibfnamefont{A.~L.} \bibnamefont{Burin}}, \bibnamefont{and}
  \bibinfo{author}{\bibfnamefont{I.~V.} \bibnamefont{Rubtsov}},
  \bibinfo{journal}{Acc. Chem. Res.} \textbf{\bibinfo{volume}{48}},
  \bibinfo{pages}{2547} (\bibinfo{year}{2015}{\natexlab{a}}).

\bibitem[{\citenamefont{Rubtsova
  et~al.}(2015{\natexlab{b}})\citenamefont{Rubtsova, Nyby, Zhang, Zhang, Zhou,
  Jayawickramarajah, Burin, and Rubtsov}}]{Rubtsova2015}
\bibinfo{author}{\bibfnamefont{N.~I.} \bibnamefont{Rubtsova}},
  \bibinfo{author}{\bibfnamefont{C.~M.} \bibnamefont{Nyby}},
  \bibinfo{author}{\bibfnamefont{H.}~\bibnamefont{Zhang}},
  \bibinfo{author}{\bibfnamefont{B.}~\bibnamefont{Zhang}},
  \bibinfo{author}{\bibfnamefont{X.}~\bibnamefont{Zhou}},
  \bibinfo{author}{\bibfnamefont{J.}~\bibnamefont{Jayawickramarajah}},
  \bibinfo{author}{\bibfnamefont{A.~L.} \bibnamefont{Burin}}, \bibnamefont{and}
  \bibinfo{author}{\bibfnamefont{I.~V.} \bibnamefont{Rubtsov}},
  \bibinfo{journal}{J. Chem. Phys.} \textbf{\bibinfo{volume}{142}},
  \bibinfo{pages}{212412} (\bibinfo{year}{2015}{\natexlab{b}}).

\bibitem[{\citenamefont{Segal et~al.}(2003)\citenamefont{Segal, Nitzan, and
  H\"anggi}}]{Nitzan2003thermal}
\bibinfo{author}{\bibfnamefont{D.}~\bibnamefont{Segal}},
  \bibinfo{author}{\bibfnamefont{A.}~\bibnamefont{Nitzan}}, \bibnamefont{and}
  \bibinfo{author}{\bibfnamefont{P.}~\bibnamefont{H\"anggi}},
  \bibinfo{journal}{J. Chem. Phys.} \textbf{\bibinfo{volume}{119}},
  \bibinfo{pages}{6840} (\bibinfo{year}{2003}).

\bibitem[{\citenamefont{Mensah et~al.}(2004)\citenamefont{Mensah, Nkrumah,
  Mensah, and Allotey}}]{Mensah2004}
\bibinfo{author}{\bibfnamefont{N.}~\bibnamefont{Mensah}},
  \bibinfo{author}{\bibfnamefont{G.}~\bibnamefont{Nkrumah}},
  \bibinfo{author}{\bibfnamefont{S.}~\bibnamefont{Mensah}}, \bibnamefont{and}
  \bibinfo{author}{\bibfnamefont{F.}~\bibnamefont{Allotey}},
  \bibinfo{journal}{Phys. Lett. A} \textbf{\bibinfo{volume}{329}},
  \bibinfo{pages}{369 } (\bibinfo{year}{2004}), ISSN \bibinfo{issn}{0375-9601}.

\bibitem[{\citenamefont{Marconnet et~al.}(2013)\citenamefont{Marconnet, Panzer,
  and Goodson}}]{Marconnet2013}
\bibinfo{author}{\bibfnamefont{A.~M.} \bibnamefont{Marconnet}},
  \bibinfo{author}{\bibfnamefont{M.~A.} \bibnamefont{Panzer}},
  \bibnamefont{and} \bibinfo{author}{\bibfnamefont{K.~E.}
  \bibnamefont{Goodson}}, \bibinfo{journal}{Rev. Mod. Phys.}
  \textbf{\bibinfo{volume}{85}}, \bibinfo{pages}{1295} (\bibinfo{year}{2013}).

\bibitem[{\citenamefont{Al-Ghalith et~al.}(2016)\citenamefont{Al-Ghalith, Ni,
  and Dumitrica}}]{AlGhalith2016}
\bibinfo{author}{\bibfnamefont{J.}~\bibnamefont{Al-Ghalith}},
  \bibinfo{author}{\bibfnamefont{Y.}~\bibnamefont{Ni}}, \bibnamefont{and}
  \bibinfo{author}{\bibfnamefont{T.}~\bibnamefont{Dumitrica}},
  \bibinfo{journal}{Phys. Chem. Chem. Phys.} \textbf{\bibinfo{volume}{18}},
  \bibinfo{pages}{9888} (\bibinfo{year}{2016}).

\bibitem[{\citenamefont{Chen  et~al.}(2005)\citenamefont{Chen , Zwolak, , and
  Ventra}}]{Chen2005}
\bibinfo{author}{\bibfnamefont{Y.-C.} \bibnamefont{Chen }},
  \bibinfo{author}{\bibfnamefont{M.}~\bibnamefont{Zwolak}}, , \bibnamefont{and}
  \bibinfo{author}{\bibfnamefont{M.~D.} \bibnamefont{Ventra}},
  \bibinfo{journal}{Nano Lett.} \textbf{\bibinfo{volume}{5}},
  \bibinfo{pages}{621} (\bibinfo{year}{2005}).

\bibitem[{\citenamefont{Losego et~al.}(2012)\citenamefont{Losego, Grady,
  Sottos, Cahill, and Braun}}]{Losego2012}
\bibinfo{author}{\bibfnamefont{M.~D.} \bibnamefont{Losego}},
  \bibinfo{author}{\bibfnamefont{M.~E.} \bibnamefont{Grady}},
  \bibinfo{author}{\bibfnamefont{N.~R.} \bibnamefont{Sottos}},
  \bibinfo{author}{\bibfnamefont{D.~G.} \bibnamefont{Cahill}},
  \bibnamefont{and} \bibinfo{author}{\bibfnamefont{P.~V.} \bibnamefont{Braun}},
  \bibinfo{journal}{Nature Mater.} \textbf{\bibinfo{volume}{11}},
  \bibinfo{pages}{502} (\bibinfo{year}{2012}).

\bibitem[{\citenamefont{O'Brien et~al.}(2013)\citenamefont{O'Brien, Shenogin,
  Liu, Chow, Laurencin, Mutin, Yamaguchi, Keblinski, and
  Ramanath}}]{OBrien2013}
\bibinfo{author}{\bibfnamefont{P.~J.} \bibnamefont{O'Brien}},
  \bibinfo{author}{\bibfnamefont{S.}~\bibnamefont{Shenogin}},
  \bibinfo{author}{\bibfnamefont{J.}~\bibnamefont{Liu}},
  \bibinfo{author}{\bibfnamefont{P.~K.} \bibnamefont{Chow}},
  \bibinfo{author}{\bibfnamefont{D.}~\bibnamefont{Laurencin}},
  \bibinfo{author}{\bibfnamefont{P.~H.} \bibnamefont{Mutin}},
  \bibinfo{author}{\bibfnamefont{M.}~\bibnamefont{Yamaguchi}},
  \bibinfo{author}{\bibfnamefont{P.}~\bibnamefont{Keblinski}},
  \bibnamefont{and} \bibinfo{author}{\bibfnamefont{G.}~\bibnamefont{Ramanath}},
  \bibinfo{journal}{Nature Mater.} \textbf{\bibinfo{volume}{12}},
  \bibinfo{pages}{118} (\bibinfo{year}{2013}).

\bibitem[{\citenamefont{Rubtsova et~al.}(2014)\citenamefont{Rubtsova, Kurnosov,
  Burin, and Rubtsov}}]{Rubtsova2014}
\bibinfo{author}{\bibfnamefont{N.~I.} \bibnamefont{Rubtsova}},
  \bibinfo{author}{\bibfnamefont{A.~A.} \bibnamefont{Kurnosov}},
  \bibinfo{author}{\bibfnamefont{A.~L.} \bibnamefont{Burin}}, \bibnamefont{and}
  \bibinfo{author}{\bibfnamefont{I.~V.} \bibnamefont{Rubtsov}},
  \bibinfo{journal}{J. Phys. Chem. B} \textbf{\bibinfo{volume}{118}},
  \bibinfo{pages}{8381} (\bibinfo{year}{2014}).

\bibitem[{\citenamefont{Segal}(2005)}]{Segal2005}
\bibinfo{author}{\bibfnamefont{D.}~\bibnamefont{Segal}},
  \bibinfo{journal}{Phys. Rev. B} \textbf{\bibinfo{volume}{72}},
  \bibinfo{pages}{165426} (\bibinfo{year}{2005}).

\bibitem[{\citenamefont{Wu and Li}(2007)}]{Wu2007}
\bibinfo{author}{\bibfnamefont{G.}~\bibnamefont{Wu}} \bibnamefont{and}
  \bibinfo{author}{\bibfnamefont{B.}~\bibnamefont{Li}}, \bibinfo{journal}{Phys.
  Rev. B} \textbf{\bibinfo{volume}{76}}, \bibinfo{pages}{085424}
  (\bibinfo{year}{2007}).

\bibitem[{\citenamefont{Wu and Segal}(2009)}]{Wu2009}
\bibinfo{author}{\bibfnamefont{L.-A.} \bibnamefont{Wu}} \bibnamefont{and}
  \bibinfo{author}{\bibfnamefont{D.}~\bibnamefont{Segal}},
  \bibinfo{journal}{Phys. Rev. Lett.} \textbf{\bibinfo{volume}{102}},
  \bibinfo{pages}{095503} (\bibinfo{year}{2009}).

\bibitem[{\citenamefont{Nicolin and
  Segal}(2011{\natexlab{a}})}]{Nicolin2011prb}
\bibinfo{author}{\bibfnamefont{L.}~\bibnamefont{Nicolin}} \bibnamefont{and}
  \bibinfo{author}{\bibfnamefont{D.}~\bibnamefont{Segal}},
  \bibinfo{journal}{Phys. Rev. B} \textbf{\bibinfo{volume}{84}},
  \bibinfo{pages}{161414} (\bibinfo{year}{2011}{\natexlab{a}}).

\bibitem[{\citenamefont{Nicolin and Segal}(2011{\natexlab{b}})}]{Nicolin2011}
\bibinfo{author}{\bibfnamefont{L.}~\bibnamefont{Nicolin}} \bibnamefont{and}
  \bibinfo{author}{\bibfnamefont{D.}~\bibnamefont{Segal}}, \bibinfo{journal}{J.
  Chem. Phys.} \textbf{\bibinfo{volume}{135}}, \bibinfo{pages}{164106}
  (\bibinfo{year}{2011}{\natexlab{b}}).

\bibitem[{\citenamefont{Gomez-Solano et~al.}(2011)\citenamefont{Gomez-Solano,
  Petrosyan, and Ciliberto}}]{GomezSolano2011}
\bibinfo{author}{\bibfnamefont{J.~R.} \bibnamefont{Gomez-Solano}},
  \bibinfo{author}{\bibfnamefont{A.}~\bibnamefont{Petrosyan}},
  \bibnamefont{and}
  \bibinfo{author}{\bibfnamefont{S.}~\bibnamefont{Ciliberto}},
  \bibinfo{journal}{Phys. Rev. Lett.} \textbf{\bibinfo{volume}{106}},
  \bibinfo{pages}{200602} (\bibinfo{year}{2011}).

\bibitem[{\citenamefont{Agarwalla et~al.}(2012)\citenamefont{Agarwalla, Li, and
  Wang}}]{Agarwalla2012}
\bibinfo{author}{\bibfnamefont{B.~K.} \bibnamefont{Agarwalla}},
  \bibinfo{author}{\bibfnamefont{B.}~\bibnamefont{Li}}, \bibnamefont{and}
  \bibinfo{author}{\bibfnamefont{J.-S.} \bibnamefont{Wang}},
  \bibinfo{journal}{Phys. Rev. E} \textbf{\bibinfo{volume}{85}},
  \bibinfo{pages}{051142} (\bibinfo{year}{2012}).

\bibitem[{\citenamefont{Arrachea et~al.}(2014)\citenamefont{Arrachea, Bode, and
  von Oppen}}]{Arrachea2014}
\bibinfo{author}{\bibfnamefont{L.}~\bibnamefont{Arrachea}},
  \bibinfo{author}{\bibfnamefont{N.}~\bibnamefont{Bode}}, \bibnamefont{and}
  \bibinfo{author}{\bibfnamefont{F.}~\bibnamefont{von Oppen}},
  \bibinfo{journal}{Phys. Rev. B} \textbf{\bibinfo{volume}{90}},
  \bibinfo{pages}{125450} (\bibinfo{year}{2014}).

\bibitem[{\citenamefont{Li et~al.}(2015)\citenamefont{Li, Duchemin, Xiong,
  Solomon, and Donadio}}]{Donadio2015}
\bibinfo{author}{\bibfnamefont{Q.}~\bibnamefont{Li}},
  \bibinfo{author}{\bibfnamefont{I.}~\bibnamefont{Duchemin}},
  \bibinfo{author}{\bibfnamefont{S.}~\bibnamefont{Xiong}},
  \bibinfo{author}{\bibfnamefont{G.~C.} \bibnamefont{Solomon}},
  \bibnamefont{and} \bibinfo{author}{\bibfnamefont{D.}~\bibnamefont{Donadio}},
  \bibinfo{journal}{J. Phys. Chem. C} \textbf{\bibinfo{volume}{119}},
  \bibinfo{pages}{24636} (\bibinfo{year}{2015}).

\bibitem[{\citenamefont{Huang et~al.}(2006)\citenamefont{Huang, Xu, Chen,
  Ventra, , and Tao}}]{Huang2006}
\bibinfo{author}{\bibfnamefont{Z.}~\bibnamefont{Huang}},
  \bibinfo{author}{\bibfnamefont{B.}~\bibnamefont{Xu}},
  \bibinfo{author}{\bibfnamefont{Y.}~\bibnamefont{Chen}},
  \bibinfo{author}{\bibfnamefont{M.~D.} \bibnamefont{Ventra}}, ,
  \bibnamefont{and} \bibinfo{author}{\bibfnamefont{N.}~\bibnamefont{Tao}},
  \bibinfo{journal}{Nano Lett.} \textbf{\bibinfo{volume}{6}},
  \bibinfo{pages}{1240} (\bibinfo{year}{2006}).

\bibitem[{\citenamefont{Tsutsui et~al.}(2008)\citenamefont{Tsutsui, Taniguchi,
  and Kawai}}]{Tsutsui2008}
\bibinfo{author}{\bibfnamefont{M.}~\bibnamefont{Tsutsui}},
  \bibinfo{author}{\bibfnamefont{M.}~\bibnamefont{Taniguchi}},
  \bibnamefont{and} \bibinfo{author}{\bibfnamefont{T.}~\bibnamefont{Kawai}},
  \bibinfo{journal}{Nano Lett.} \textbf{\bibinfo{volume}{8}},
  \bibinfo{pages}{3293} (\bibinfo{year}{2008}).

\bibitem[{\citenamefont{Hoffmann et~al.}(2009)\citenamefont{Hoffmann, Nilsson,
  Matthews, Nakpathomkun, Persson, Samuelson, and Linke}}]{Hoffman2009}
\bibinfo{author}{\bibfnamefont{E.~A.} \bibnamefont{Hoffmann}},
  \bibinfo{author}{\bibfnamefont{H.~A.} \bibnamefont{Nilsson}},
  \bibinfo{author}{\bibfnamefont{J.~E.} \bibnamefont{Matthews}},
  \bibinfo{author}{\bibfnamefont{N.}~\bibnamefont{Nakpathomkun}},
  \bibinfo{author}{\bibfnamefont{A.~I.} \bibnamefont{Persson}},
  \bibinfo{author}{\bibfnamefont{L.}~\bibnamefont{Samuelson}},
  \bibnamefont{and} \bibinfo{author}{\bibfnamefont{H.}~\bibnamefont{Linke}},
  \bibinfo{journal}{Nano Lett.} \textbf{\bibinfo{volume}{9}},
  \bibinfo{pages}{779} (\bibinfo{year}{2009}).

\bibitem[{\citenamefont{Chen et~al.}(2014)\citenamefont{Chen, Wheeler,
  Di~Ventra, and Natelson}}]{Chen2014}
\bibinfo{author}{\bibfnamefont{R.}~\bibnamefont{Chen}},
  \bibinfo{author}{\bibfnamefont{P.~J.} \bibnamefont{Wheeler}},
  \bibinfo{author}{\bibfnamefont{M.}~\bibnamefont{Di~Ventra}},
  \bibnamefont{and} \bibinfo{author}{\bibfnamefont{D.}~\bibnamefont{Natelson}},
  \bibinfo{journal}{Sci. Rep.} \textbf{\bibinfo{volume}{4}}
  (\bibinfo{year}{2014}).

\bibitem[{\citenamefont{Maher et~al.}(2006)\citenamefont{Maher, Cohen, Le~Ru,
  and Etchegoin}}]{Maher2006}
\bibinfo{author}{\bibfnamefont{R.~C.} \bibnamefont{Maher}},
  \bibinfo{author}{\bibfnamefont{L.~F.} \bibnamefont{Cohen}},
  \bibinfo{author}{\bibfnamefont{E.~C.} \bibnamefont{Le~Ru}}, \bibnamefont{and}
  \bibinfo{author}{\bibfnamefont{P.~G.} \bibnamefont{Etchegoin}},
  \bibinfo{journal}{Faraday Discuss.} \textbf{\bibinfo{volume}{132}},
  \bibinfo{pages}{77} (\bibinfo{year}{2006}).

\bibitem[{\citenamefont{Ioffe et~al.}(2008)\citenamefont{Ioffe, Shamai, Ophir,
  Noy, Yutsis, Kfir, Cheshnovsky, and Selzer}}]{Ioffe2008}
\bibinfo{author}{\bibfnamefont{Z.}~\bibnamefont{Ioffe}},
  \bibinfo{author}{\bibfnamefont{T.}~\bibnamefont{Shamai}},
  \bibinfo{author}{\bibfnamefont{A.}~\bibnamefont{Ophir}},
  \bibinfo{author}{\bibfnamefont{G.}~\bibnamefont{Noy}},
  \bibinfo{author}{\bibfnamefont{I.}~\bibnamefont{Yutsis}},
  \bibinfo{author}{\bibfnamefont{K.}~\bibnamefont{Kfir}},
  \bibinfo{author}{\bibfnamefont{O.}~\bibnamefont{Cheshnovsky}},
  \bibnamefont{and} \bibinfo{author}{\bibfnamefont{Y.}~\bibnamefont{Selzer}},
  \bibinfo{journal}{Nature Nanotech.} \textbf{\bibinfo{volume}{3}},
  \bibinfo{pages}{727} (\bibinfo{year}{2008}).

\bibitem[{\citenamefont{Ward et~al.}(2011)\citenamefont{Ward, Corley, Tour, and
  Natelson}}]{Ward2011}
\bibinfo{author}{\bibfnamefont{D.~R.} \bibnamefont{Ward}},
  \bibinfo{author}{\bibfnamefont{D.~A.} \bibnamefont{Corley}},
  \bibinfo{author}{\bibfnamefont{J.~M.} \bibnamefont{Tour}}, \bibnamefont{and}
  \bibinfo{author}{\bibfnamefont{D.}~\bibnamefont{Natelson}},
  \bibinfo{journal}{Nature Nanotech.} \textbf{\bibinfo{volume}{6}},
  \bibinfo{pages}{33} (\bibinfo{year}{2011}).

\bibitem[{\citenamefont{Dang et~al.}(2011)\citenamefont{Dang, Bolme, Moore, and
  McGrane}}]{Dang2011}
\bibinfo{author}{\bibfnamefont{N.~C.} \bibnamefont{Dang}},
  \bibinfo{author}{\bibfnamefont{C.~A.} \bibnamefont{Bolme}},
  \bibinfo{author}{\bibfnamefont{D.~S.} \bibnamefont{Moore}}, \bibnamefont{and}
  \bibinfo{author}{\bibfnamefont{S.~D.} \bibnamefont{McGrane}},
  \bibinfo{journal}{Phys. Rev. Lett.} \textbf{\bibinfo{volume}{107}},
  \bibinfo{pages}{043001} (\bibinfo{year}{2011}).

\bibitem[{\citenamefont{Sadat et~al.}(2010)\citenamefont{Sadat, Tan, Chua, and
  Reddy}}]{Sadat2010}
\bibinfo{author}{\bibfnamefont{S.}~\bibnamefont{Sadat}},
  \bibinfo{author}{\bibfnamefont{A.}~\bibnamefont{Tan}},
  \bibinfo{author}{\bibfnamefont{Y.~J.} \bibnamefont{Chua}}, \bibnamefont{and}
  \bibinfo{author}{\bibfnamefont{P.}~\bibnamefont{Reddy}},
  \bibinfo{journal}{Nano Lett.} \textbf{\bibinfo{volume}{10}},
  \bibinfo{pages}{2613} (\bibinfo{year}{2010}).

\bibitem[{\citenamefont{Menges et~al.}(2012)\citenamefont{Menges, Riel,
  Stemmer, and Gotsmann}}]{Menges2012}
\bibinfo{author}{\bibfnamefont{F.}~\bibnamefont{Menges}},
  \bibinfo{author}{\bibfnamefont{H.}~\bibnamefont{Riel}},
  \bibinfo{author}{\bibfnamefont{A.}~\bibnamefont{Stemmer}}, \bibnamefont{and}
  \bibinfo{author}{\bibfnamefont{B.}~\bibnamefont{Gotsmann}},
  \bibinfo{journal}{Nano Lett.} \textbf{\bibinfo{volume}{12}},
  \bibinfo{pages}{596} (\bibinfo{year}{2012}).

\bibitem[{\citenamefont{Lee et~al.}(2013)\citenamefont{Lee, Kim, Jeong, Zotti,
  Pauly, Cuevas, and Reddy}}]{Lee2013}
\bibinfo{author}{\bibfnamefont{W.}~\bibnamefont{Lee}},
  \bibinfo{author}{\bibfnamefont{K.}~\bibnamefont{Kim}},
  \bibinfo{author}{\bibfnamefont{W.}~\bibnamefont{Jeong}},
  \bibinfo{author}{\bibfnamefont{L.~A.} \bibnamefont{Zotti}},
  \bibinfo{author}{\bibfnamefont{F.}~\bibnamefont{Pauly}},
  \bibinfo{author}{\bibfnamefont{J.~C.} \bibnamefont{Cuevas}},
  \bibnamefont{and} \bibinfo{author}{\bibfnamefont{P.}~\bibnamefont{Reddy}},
  \bibinfo{journal}{Nature} \textbf{\bibinfo{volume}{498}},
  \bibinfo{pages}{209} (\bibinfo{year}{2013}).

\bibitem[{\citenamefont{Desiatov et~al.}(2014)\citenamefont{Desiatov, Goykhman,
  and Levy}}]{Desiatov2014}
\bibinfo{author}{\bibfnamefont{B.}~\bibnamefont{Desiatov}},
  \bibinfo{author}{\bibfnamefont{I.}~\bibnamefont{Goykhman}}, \bibnamefont{and}
  \bibinfo{author}{\bibfnamefont{U.}~\bibnamefont{Levy}},
  \bibinfo{journal}{Nano Lett.} \textbf{\bibinfo{volume}{14}},
  \bibinfo{pages}{648} (\bibinfo{year}{2014}).

\bibitem[{\citenamefont{Chen et~al.}(2015)\citenamefont{Chen, Shan, Guan, Wang,
  Zhu, and Tao}}]{Chen2015}
\bibinfo{author}{\bibfnamefont{Z.}~\bibnamefont{Chen}},
  \bibinfo{author}{\bibfnamefont{X.}~\bibnamefont{Shan}},
  \bibinfo{author}{\bibfnamefont{Y.}~\bibnamefont{Guan}},
  \bibinfo{author}{\bibfnamefont{S.}~\bibnamefont{Wang}},
  \bibinfo{author}{\bibfnamefont{J.-J.} \bibnamefont{Zhu}}, \bibnamefont{and}
  \bibinfo{author}{\bibfnamefont{N.}~\bibnamefont{Tao}}, \bibinfo{journal}{ACS
  Nano} \textbf{\bibinfo{volume}{9}}, \bibinfo{pages}{11574}
  (\bibinfo{year}{2015}).

\bibitem[{\citenamefont{Hu et~al.}(2015)\citenamefont{Hu, Zeng, Minnich,
  Dresselhaus, and Chen}}]{Hu2015}
\bibinfo{author}{\bibfnamefont{Y.}~\bibnamefont{Hu}},
  \bibinfo{author}{\bibfnamefont{L.}~\bibnamefont{Zeng}},
  \bibinfo{author}{\bibfnamefont{A.~J.} \bibnamefont{Minnich}},
  \bibinfo{author}{\bibfnamefont{M.~S.} \bibnamefont{Dresselhaus}},
  \bibnamefont{and} \bibinfo{author}{\bibfnamefont{G.}~\bibnamefont{Chen}},
  \bibinfo{journal}{Nature Nanotech.} \textbf{\bibinfo{volume}{10}},
  \bibinfo{pages}{701} (\bibinfo{year}{2015}).

\bibitem[{\citenamefont{Mecklenburg et~al.}(2015)\citenamefont{Mecklenburg,
  Hubbard, White, Dhall, Cronin, Aloni, and Regan}}]{Mecklenburg2015}
\bibinfo{author}{\bibfnamefont{M.}~\bibnamefont{Mecklenburg}},
  \bibinfo{author}{\bibfnamefont{W.~A.} \bibnamefont{Hubbard}},
  \bibinfo{author}{\bibfnamefont{E.~R.} \bibnamefont{White}},
  \bibinfo{author}{\bibfnamefont{R.}~\bibnamefont{Dhall}},
  \bibinfo{author}{\bibfnamefont{S.~B.} \bibnamefont{Cronin}},
  \bibinfo{author}{\bibfnamefont{S.}~\bibnamefont{Aloni}}, \bibnamefont{and}
  \bibinfo{author}{\bibfnamefont{B.~C.} \bibnamefont{Regan}},
  \bibinfo{journal}{Science} \textbf{\bibinfo{volume}{347}},
  \bibinfo{pages}{629} (\bibinfo{year}{2015}).

\bibitem[{\citenamefont{Schwarzer et~al.}(2004)\citenamefont{Schwarzer, Kutne,
  Schr{\"o}der, and Troe}}]{Schwarzer2004}
\bibinfo{author}{\bibfnamefont{D.}~\bibnamefont{Schwarzer}},
  \bibinfo{author}{\bibfnamefont{P.}~\bibnamefont{Kutne}},
  \bibinfo{author}{\bibfnamefont{C.}~\bibnamefont{Schr{\"o}der}},
  \bibnamefont{and} \bibinfo{author}{\bibfnamefont{J.}~\bibnamefont{Troe}},
  \bibinfo{journal}{J. Chem. Phys.} \textbf{\bibinfo{volume}{121}},
  \bibinfo{pages}{1754} (\bibinfo{year}{2004}).

\bibitem[{\citenamefont{Wang et~al.}(2007)\citenamefont{Wang, Carter,
  Lagutchev, Koh, Seong, Cahill, and Dlott}}]{Wang2007}
\bibinfo{author}{\bibfnamefont{Z.}~\bibnamefont{Wang}},
  \bibinfo{author}{\bibfnamefont{J.~A.} \bibnamefont{Carter}},
  \bibinfo{author}{\bibfnamefont{A.}~\bibnamefont{Lagutchev}},
  \bibinfo{author}{\bibfnamefont{Y.~K.} \bibnamefont{Koh}},
  \bibinfo{author}{\bibfnamefont{N.-H.} \bibnamefont{Seong}},
  \bibinfo{author}{\bibfnamefont{D.~G.} \bibnamefont{Cahill}},
  \bibnamefont{and} \bibinfo{author}{\bibfnamefont{D.~D.} \bibnamefont{Dlott}},
  \bibinfo{journal}{Science} \textbf{\bibinfo{volume}{317}},
  \bibinfo{pages}{787} (\bibinfo{year}{2007}).

\bibitem[{\citenamefont{Carter et~al.}(2008)\citenamefont{Carter, Wang, , and
  Dlott}}]{Carter2008}
\bibinfo{author}{\bibfnamefont{J.~A.} \bibnamefont{Carter}},
  \bibinfo{author}{\bibfnamefont{Z.}~\bibnamefont{Wang}}, , \bibnamefont{and}
  \bibinfo{author}{\bibfnamefont{D.~D.} \bibnamefont{Dlott}},
  \bibinfo{journal}{J. Phys. Chem. A} \textbf{\bibinfo{volume}{112}},
  \bibinfo{pages}{3523} (\bibinfo{year}{2008}).

\bibitem[{\citenamefont{Wang et~al.}(2008)\citenamefont{Wang, Cahill, Carter,
  Koh, Lagutchev, Seong, and Dlott}}]{Wang2008}
\bibinfo{author}{\bibfnamefont{Z.}~\bibnamefont{Wang}},
  \bibinfo{author}{\bibfnamefont{D.~G.} \bibnamefont{Cahill}},
  \bibinfo{author}{\bibfnamefont{J.~A.} \bibnamefont{Carter}},
  \bibinfo{author}{\bibfnamefont{Y.~K.} \bibnamefont{Koh}},
  \bibinfo{author}{\bibfnamefont{A.}~\bibnamefont{Lagutchev}},
  \bibinfo{author}{\bibfnamefont{N.-H.} \bibnamefont{Seong}}, \bibnamefont{and}
  \bibinfo{author}{\bibfnamefont{D.~D.} \bibnamefont{Dlott}},
  \bibinfo{journal}{Computers Phys.} \textbf{\bibinfo{volume}{350}},
  \bibinfo{pages}{31 } (\bibinfo{year}{2008}).

\bibitem[{\citenamefont{Pein et~al.}(2013)\citenamefont{Pein, Sun, and
  Dlott}}]{Pein2013}
\bibinfo{author}{\bibfnamefont{B.~C.} \bibnamefont{Pein}},
  \bibinfo{author}{\bibfnamefont{Y.}~\bibnamefont{Sun}}, \bibnamefont{and}
  \bibinfo{author}{\bibfnamefont{D.~D.} \bibnamefont{Dlott}},
  \bibinfo{journal}{J. Phys. Chem. B} \textbf{\bibinfo{volume}{117}},
  \bibinfo{pages}{10898} (\bibinfo{year}{2013}).

\bibitem[{\citenamefont{Kasyanenko et~al.}(2011)\citenamefont{Kasyanenko,
  Tesar, Rubtsov, Burin, and Rubtsov}}]{Kasyanenko2011}
\bibinfo{author}{\bibfnamefont{V.~M.} \bibnamefont{Kasyanenko}},
  \bibinfo{author}{\bibfnamefont{S.~L.} \bibnamefont{Tesar}},
  \bibinfo{author}{\bibfnamefont{G.~I.} \bibnamefont{Rubtsov}},
  \bibinfo{author}{\bibfnamefont{A.~L.} \bibnamefont{Burin}}, \bibnamefont{and}
  \bibinfo{author}{\bibfnamefont{I.~V.} \bibnamefont{Rubtsov}},
  \bibinfo{journal}{J. Phys. Chem. B} \textbf{\bibinfo{volume}{115}},
  \bibinfo{pages}{11063} (\bibinfo{year}{2011}).

\bibitem[{\citenamefont{Meier et~al.}(2014)\citenamefont{Meier, Menges,
  Nirmalraj, H\"olscher, Riel, and Gotsmann}}]{Meier2014}
\bibinfo{author}{\bibfnamefont{T.}~\bibnamefont{Meier}},
  \bibinfo{author}{\bibfnamefont{F.}~\bibnamefont{Menges}},
  \bibinfo{author}{\bibfnamefont{P.}~\bibnamefont{Nirmalraj}},
  \bibinfo{author}{\bibfnamefont{H.}~\bibnamefont{H\"olscher}},
  \bibinfo{author}{\bibfnamefont{H.}~\bibnamefont{Riel}}, \bibnamefont{and}
  \bibinfo{author}{\bibfnamefont{B.}~\bibnamefont{Gotsmann}},
  \bibinfo{journal}{Phys. Rev. Lett.} \textbf{\bibinfo{volume}{113}},
  \bibinfo{pages}{060801} (\bibinfo{year}{2014}).

\bibitem[{\citenamefont{Kurnosov et~al.}(2015)\citenamefont{Kurnosov, Rubtsov,
  and Burin}}]{Kurnosov2015}
\bibinfo{author}{\bibfnamefont{A.~A.} \bibnamefont{Kurnosov}},
  \bibinfo{author}{\bibfnamefont{I.~V.} \bibnamefont{Rubtsov}},
  \bibnamefont{and} \bibinfo{author}{\bibfnamefont{A.~L.} \bibnamefont{Burin}},
  \bibinfo{journal}{J. Chem. Phys.} \textbf{\bibinfo{volume}{142}},
  \bibinfo{pages}{011101} (\bibinfo{year}{2015}).

\bibitem[{\citenamefont{Yue et~al.}(2015)\citenamefont{Yue, Qasim, Kurnosov,
  Rubtsova, Mackin, Zhang, Zhang, Zhou, Jayawickramarajah, Burin
  et~al.}}]{Yue2015}
\bibinfo{author}{\bibfnamefont{Y.}~\bibnamefont{Yue}},
  \bibinfo{author}{\bibfnamefont{L.~N.} \bibnamefont{Qasim}},
  \bibinfo{author}{\bibfnamefont{A.~A.} \bibnamefont{Kurnosov}},
  \bibinfo{author}{\bibfnamefont{N.~I.} \bibnamefont{Rubtsova}},
  \bibinfo{author}{\bibfnamefont{R.~T.} \bibnamefont{Mackin}},
  \bibinfo{author}{\bibfnamefont{H.}~\bibnamefont{Zhang}},
  \bibinfo{author}{\bibfnamefont{B.}~\bibnamefont{Zhang}},
  \bibinfo{author}{\bibfnamefont{X.}~\bibnamefont{Zhou}},
  \bibinfo{author}{\bibfnamefont{J.}~\bibnamefont{Jayawickramarajah}},
  \bibinfo{author}{\bibfnamefont{A.~L.} \bibnamefont{Burin}},
  \bibnamefont{et~al.}, \bibinfo{journal}{J. Phys. Chem. B}
  \textbf{\bibinfo{volume}{119}}, \bibinfo{pages}{6448} (\bibinfo{year}{2015}).

\bibitem[{\citenamefont{Galperin
  et~al.}(2007{\natexlab{b}})\citenamefont{Galperin, Nitzan, and
  Ratner}}]{Nitzan2007}
\bibinfo{author}{\bibfnamefont{M.}~\bibnamefont{Galperin}},
  \bibinfo{author}{\bibfnamefont{A.}~\bibnamefont{Nitzan}}, \bibnamefont{and}
  \bibinfo{author}{\bibfnamefont{M.~A.} \bibnamefont{Ratner}},
  \bibinfo{journal}{Phys. Rev. B} \textbf{\bibinfo{volume}{75}},
  \bibinfo{pages}{155312} (\bibinfo{year}{2007}{\natexlab{b}}).

\bibitem[{\citenamefont{Galperin et~al.}(2009)\citenamefont{Galperin, Saito,
  Balatsky, and Nitzan}}]{Galperin2009}
\bibinfo{author}{\bibfnamefont{M.}~\bibnamefont{Galperin}},
  \bibinfo{author}{\bibfnamefont{K.}~\bibnamefont{Saito}},
  \bibinfo{author}{\bibfnamefont{A.~V.} \bibnamefont{Balatsky}},
  \bibnamefont{and} \bibinfo{author}{\bibfnamefont{A.}~\bibnamefont{Nitzan}},
  \bibinfo{journal}{Phys. Rev. B} \textbf{\bibinfo{volume}{80}},
  \bibinfo{pages}{115427} (\bibinfo{year}{2009}).

\bibitem[{\citenamefont{Galperin and
  Nitzan}(2011{\natexlab{a}})}]{Nitzan2011jpcl}
\bibinfo{author}{\bibfnamefont{M.}~\bibnamefont{Galperin}} \bibnamefont{and}
  \bibinfo{author}{\bibfnamefont{A.}~\bibnamefont{Nitzan}},
  \bibinfo{journal}{J. Phys. Chem. Lett.} \textbf{\bibinfo{volume}{2}},
  \bibinfo{pages}{2110} (\bibinfo{year}{2011}{\natexlab{a}}).

\bibitem[{\citenamefont{Galperin and
  Nitzan}(2011{\natexlab{b}})}]{Nitzan2011prb}
\bibinfo{author}{\bibfnamefont{M.}~\bibnamefont{Galperin}} \bibnamefont{and}
  \bibinfo{author}{\bibfnamefont{A.}~\bibnamefont{Nitzan}},
  \bibinfo{journal}{Phys. Rev. B} \textbf{\bibinfo{volume}{84}},
  \bibinfo{pages}{195325} (\bibinfo{year}{2011}{\natexlab{b}}).

\bibitem[{\citenamefont{Horsfield et~al.}(2006)\citenamefont{Horsfield, Bowler,
  Ness, S\'anchez, Todorov, and Fisher}}]{Horsfield2006}
\bibinfo{author}{\bibfnamefont{A.~P.} \bibnamefont{Horsfield}},
  \bibinfo{author}{\bibfnamefont{D.~R.} \bibnamefont{Bowler}},
  \bibinfo{author}{\bibfnamefont{H.}~\bibnamefont{Ness}},
  \bibinfo{author}{\bibfnamefont{C.~G.} \bibnamefont{S\'anchez}},
  \bibinfo{author}{\bibfnamefont{T.~N.} \bibnamefont{Todorov}},
  \bibnamefont{and} \bibinfo{author}{\bibfnamefont{A.~J.}
  \bibnamefont{Fisher}}, \bibinfo{journal}{Rep. Prog. Phys.}
  \textbf{\bibinfo{volume}{69}}, \bibinfo{pages}{1195} (\bibinfo{year}{2006}).

\bibitem[{\citenamefont{D'Agosta and Ventra}(2008)}]{DAgosta2008}
\bibinfo{author}{\bibfnamefont{R.}~\bibnamefont{D'Agosta}} \bibnamefont{and}
  \bibinfo{author}{\bibfnamefont{M.~D.} \bibnamefont{Ventra}},
  \bibinfo{journal}{J. Phys.: Condens. Matter} \textbf{\bibinfo{volume}{20}},
  \bibinfo{pages}{374102} (\bibinfo{year}{2008}).

\bibitem[{\citenamefont{Asai}(2011)}]{Asai2011}
\bibinfo{author}{\bibfnamefont{Y.}~\bibnamefont{Asai}}, \bibinfo{journal}{Phys.
  Rev. B} \textbf{\bibinfo{volume}{84}}, \bibinfo{pages}{085436}
  (\bibinfo{year}{2011}).

\bibitem[{\citenamefont{Asai}(2015)}]{Asai2015}
\bibinfo{author}{\bibfnamefont{Y.}~\bibnamefont{Asai}}, \bibinfo{journal}{Phys.
  Rev. B} \textbf{\bibinfo{volume}{91}}, \bibinfo{pages}{161402}
  (\bibinfo{year}{2015}).

\bibitem[{\citenamefont{Reddy et~al.}(2007)\citenamefont{Reddy, Jang, Segalman,
  and Majumdar}}]{Reddy2007}
\bibinfo{author}{\bibfnamefont{P.}~\bibnamefont{Reddy}},
  \bibinfo{author}{\bibfnamefont{S.-Y.} \bibnamefont{Jang}},
  \bibinfo{author}{\bibfnamefont{R.~A.} \bibnamefont{Segalman}},
  \bibnamefont{and} \bibinfo{author}{\bibfnamefont{A.}~\bibnamefont{Majumdar}},
  \bibinfo{journal}{Science} \textbf{\bibinfo{volume}{315}},
  \bibinfo{pages}{1568} (\bibinfo{year}{2007}).

\bibitem[{\citenamefont{Malen et~al.}(2009)\citenamefont{Malen, Doak, Baheti,
  Tilley, Majumdar, and Segalman}}]{Malen2009}
\bibinfo{author}{\bibfnamefont{J.~A.} \bibnamefont{Malen}},
  \bibinfo{author}{\bibfnamefont{P.}~\bibnamefont{Doak}},
  \bibinfo{author}{\bibfnamefont{K.}~\bibnamefont{Baheti}},
  \bibinfo{author}{\bibfnamefont{T.~D.} \bibnamefont{Tilley}},
  \bibinfo{author}{\bibfnamefont{A.}~\bibnamefont{Majumdar}}, \bibnamefont{and}
  \bibinfo{author}{\bibfnamefont{R.~A.} \bibnamefont{Segalman}},
  \bibinfo{journal}{Nano Lett.} \textbf{\bibinfo{volume}{9}},
  \bibinfo{pages}{3406} (\bibinfo{year}{2009}).

\bibitem[{\citenamefont{Malen et~al.}(2010)\citenamefont{Malen, Yee, Majumdar,
  and Segalman}}]{Malen2010}
\bibinfo{author}{\bibfnamefont{J.~A.} \bibnamefont{Malen}},
  \bibinfo{author}{\bibfnamefont{S.~K.} \bibnamefont{Yee}},
  \bibinfo{author}{\bibfnamefont{A.}~\bibnamefont{Majumdar}}, \bibnamefont{and}
  \bibinfo{author}{\bibfnamefont{R.~A.} \bibnamefont{Segalman}},
  \bibinfo{journal}{Chem. Phys. Lett.} \textbf{\bibinfo{volume}{491}},
  \bibinfo{pages}{109} (\bibinfo{year}{2010}).

\bibitem[{\citenamefont{Tan et~al.}(2011)\citenamefont{Tan, Balachandran,
  Sadat, Gavini, Dunietz, Jang, and Reddy}}]{Tan2011}
\bibinfo{author}{\bibfnamefont{A.}~\bibnamefont{Tan}},
  \bibinfo{author}{\bibfnamefont{J.}~\bibnamefont{Balachandran}},
  \bibinfo{author}{\bibfnamefont{S.}~\bibnamefont{Sadat}},
  \bibinfo{author}{\bibfnamefont{V.}~\bibnamefont{Gavini}},
  \bibinfo{author}{\bibfnamefont{B.~D.} \bibnamefont{Dunietz}},
  \bibinfo{author}{\bibfnamefont{S.-Y.} \bibnamefont{Jang}}, \bibnamefont{and}
  \bibinfo{author}{\bibfnamefont{P.}~\bibnamefont{Reddy}}, \bibinfo{journal}{J.
  Am. Chem. Soc.} \textbf{\bibinfo{volume}{133}}, \bibinfo{pages}{8838}
  (\bibinfo{year}{2011}).

\bibitem[{\citenamefont{Kim et~al.}(2014)\citenamefont{Kim, Jeong, Kim, Lee,
  and Reddy}}]{Kim2014}
\bibinfo{author}{\bibfnamefont{Y.}~\bibnamefont{Kim}},
  \bibinfo{author}{\bibfnamefont{W.}~\bibnamefont{Jeong}},
  \bibinfo{author}{\bibfnamefont{K.}~\bibnamefont{Kim}},
  \bibinfo{author}{\bibfnamefont{W.}~\bibnamefont{Lee}}, \bibnamefont{and}
  \bibinfo{author}{\bibfnamefont{P.}~\bibnamefont{Reddy}},
  \bibinfo{journal}{Nature Nanotech.} \textbf{\bibinfo{volume}{9}},
  \bibinfo{pages}{881} (\bibinfo{year}{2014}).

\bibitem[{\citenamefont{Paulsson and Datta}(2003)}]{Paulsson2003}
\bibinfo{author}{\bibfnamefont{M.}~\bibnamefont{Paulsson}} \bibnamefont{and}
  \bibinfo{author}{\bibfnamefont{S.}~\bibnamefont{Datta}},
  \bibinfo{journal}{Phys. Rev. B} \textbf{\bibinfo{volume}{67}},
  \bibinfo{pages}{241403} (\bibinfo{year}{2003}).

\bibitem[{\citenamefont{Koch et~al.}(2004)\citenamefont{Koch, von Oppen, Oreg,
  and Sela}}]{Koch2004}
\bibinfo{author}{\bibfnamefont{J.}~\bibnamefont{Koch}},
  \bibinfo{author}{\bibfnamefont{F.}~\bibnamefont{von Oppen}},
  \bibinfo{author}{\bibfnamefont{Y.}~\bibnamefont{Oreg}}, \bibnamefont{and}
  \bibinfo{author}{\bibfnamefont{E.}~\bibnamefont{Sela}},
  \bibinfo{journal}{Phys. Rev. B} \textbf{\bibinfo{volume}{70}},
  \bibinfo{pages}{195107} (\bibinfo{year}{2004}).

\bibitem[{\citenamefont{Pauly et~al.}(2008)\citenamefont{Pauly, Viljas, and
  Cuevas}}]{Pauly2008}
\bibinfo{author}{\bibfnamefont{F.}~\bibnamefont{Pauly}},
  \bibinfo{author}{\bibfnamefont{J.~K.} \bibnamefont{Viljas}},
  \bibnamefont{and} \bibinfo{author}{\bibfnamefont{J.~C.}
  \bibnamefont{Cuevas}}, \bibinfo{journal}{Phys. Rev. B}
  \textbf{\bibinfo{volume}{78}}, \bibinfo{pages}{035315}
  (\bibinfo{year}{2008}).

\bibitem[{\citenamefont{Bergfield and Stafford}(2009)}]{Bergfield2009}
\bibinfo{author}{\bibfnamefont{J.~P.} \bibnamefont{Bergfield}}
  \bibnamefont{and} \bibinfo{author}{\bibfnamefont{C.~A.}
  \bibnamefont{Stafford}}, \bibinfo{journal}{Nano Lett.}
  \textbf{\bibinfo{volume}{9}}, \bibinfo{pages}{3072} (\bibinfo{year}{2009}).

\bibitem[{\citenamefont{Bergfield et~al.}(2010)\citenamefont{Bergfield, Solis,
  and Stafford}}]{Bergfield2010}
\bibinfo{author}{\bibfnamefont{J.~P.} \bibnamefont{Bergfield}},
  \bibinfo{author}{\bibfnamefont{M.~A.} \bibnamefont{Solis}}, \bibnamefont{and}
  \bibinfo{author}{\bibfnamefont{C.~A.} \bibnamefont{Stafford}},
  \bibinfo{journal}{ACS Nano} \textbf{\bibinfo{volume}{4}},
  \bibinfo{pages}{5314} (\bibinfo{year}{2010}).

\bibitem[{\citenamefont{Ke et~al.}(2009)\citenamefont{Ke, Yang, Curtarolo, and
  Baranger}}]{Ke2009}
\bibinfo{author}{\bibfnamefont{S.-H.} \bibnamefont{Ke}},
  \bibinfo{author}{\bibfnamefont{W.}~\bibnamefont{Yang}},
  \bibinfo{author}{\bibfnamefont{S.}~\bibnamefont{Curtarolo}},
  \bibnamefont{and} \bibinfo{author}{\bibfnamefont{H.~U.}
  \bibnamefont{Baranger}}, \bibinfo{journal}{Nano Lett.}
  \textbf{\bibinfo{volume}{9}}, \bibinfo{pages}{1011} (\bibinfo{year}{2009}).

\bibitem[{\citenamefont{Liu and Chen}(2009)}]{Liu2009}
\bibinfo{author}{\bibfnamefont{Y.-S.} \bibnamefont{Liu}} \bibnamefont{and}
  \bibinfo{author}{\bibfnamefont{Y.-C.} \bibnamefont{Chen}},
  \bibinfo{journal}{Phys. Rev. B} \textbf{\bibinfo{volume}{79}},
  \bibinfo{pages}{193101} (\bibinfo{year}{2009}).

\bibitem[{\citenamefont{Ren et~al.}(2012)\citenamefont{Ren, Zhu, Gubernatis,
  Wang, and Li}}]{Ren2012}
\bibinfo{author}{\bibfnamefont{J.}~\bibnamefont{Ren}},
  \bibinfo{author}{\bibfnamefont{J.-X.} \bibnamefont{Zhu}},
  \bibinfo{author}{\bibfnamefont{J.~E.} \bibnamefont{Gubernatis}},
  \bibinfo{author}{\bibfnamefont{C.}~\bibnamefont{Wang}}, \bibnamefont{and}
  \bibinfo{author}{\bibfnamefont{B.}~\bibnamefont{Li}}, \bibinfo{journal}{Phys.
  Rev. B} \textbf{\bibinfo{volume}{85}}, \bibinfo{pages}{155443}
  (\bibinfo{year}{2012}).

\bibitem[{\citenamefont{Wang et~al.}(2011)\citenamefont{Wang, Zhou, and
  Yang}}]{Wang2011}
\bibinfo{author}{\bibfnamefont{Y.}~\bibnamefont{Wang}},
  \bibinfo{author}{\bibfnamefont{J.}~\bibnamefont{Zhou}}, \bibnamefont{and}
  \bibinfo{author}{\bibfnamefont{R.}~\bibnamefont{Yang}}, \bibinfo{journal}{J.
  Phys. Chem. C} \textbf{\bibinfo{volume}{115}}, \bibinfo{pages}{24418}
  (\bibinfo{year}{2011}).

\bibitem[{\citenamefont{Lee et~al.}(2014)\citenamefont{Lee, Cho, Lyeo, and
  Kim}}]{Lee2014}
\bibinfo{author}{\bibfnamefont{E.-S.} \bibnamefont{Lee}},
  \bibinfo{author}{\bibfnamefont{S.}~\bibnamefont{Cho}},
  \bibinfo{author}{\bibfnamefont{H.-K.} \bibnamefont{Lyeo}}, \bibnamefont{and}
  \bibinfo{author}{\bibfnamefont{Y.-H.} \bibnamefont{Kim}},
  \bibinfo{journal}{Phys. Rev. Lett.} \textbf{\bibinfo{volume}{112}},
  \bibinfo{pages}{136601} (\bibinfo{year}{2014}).

\bibitem[{\citenamefont{Amanatidis et~al.}(2015)\citenamefont{Amanatidis, Kao,
  Du, Pao, and Chen}}]{Amanatidis2015}
\bibinfo{author}{\bibfnamefont{I.}~\bibnamefont{Amanatidis}},
  \bibinfo{author}{\bibfnamefont{J.-Y.} \bibnamefont{Kao}},
  \bibinfo{author}{\bibfnamefont{L.-Y.} \bibnamefont{Du}},
  \bibinfo{author}{\bibfnamefont{C.-W.} \bibnamefont{Pao}}, \bibnamefont{and}
  \bibinfo{author}{\bibfnamefont{Y.-C.} \bibnamefont{Chen}},
  \bibinfo{journal}{J. Phys. Chem. C} \textbf{\bibinfo{volume}{119}},
  \bibinfo{pages}{28728} (\bibinfo{year}{2015}).

\bibitem[{\citenamefont{Simine et~al.}(2015)\citenamefont{Simine, Chen, and
  Segal}}]{Simine2015}
\bibinfo{author}{\bibfnamefont{L.}~\bibnamefont{Simine}},
  \bibinfo{author}{\bibfnamefont{W.~J.} \bibnamefont{Chen}}, \bibnamefont{and}
  \bibinfo{author}{\bibfnamefont{D.}~\bibnamefont{Segal}}, \bibinfo{journal}{J.
  Phys. Chem. C} \textbf{\bibinfo{volume}{119}}, \bibinfo{pages}{12097}
  (\bibinfo{year}{2015}).

\bibitem[{\citenamefont{Walczak}(2007)}]{Walczak2007}
\bibinfo{author}{\bibfnamefont{K.}~\bibnamefont{Walczak}},
  \bibinfo{journal}{Physica B} \textbf{\bibinfo{volume}{392}},
  \bibinfo{pages}{173 } (\bibinfo{year}{2007}), ISSN \bibinfo{issn}{0921-4526}.

\bibitem[{\citenamefont{Koch et~al.}(2014)\citenamefont{Koch, Loos, and
  Fehske}}]{Koch2014}
\bibinfo{author}{\bibfnamefont{T.}~\bibnamefont{Koch}},
  \bibinfo{author}{\bibfnamefont{J.}~\bibnamefont{Loos}}, \bibnamefont{and}
  \bibinfo{author}{\bibfnamefont{H.}~\bibnamefont{Fehske}},
  \bibinfo{journal}{Phys. Rev. B} \textbf{\bibinfo{volume}{89}},
  \bibinfo{pages}{155133} (\bibinfo{year}{2014}).

\bibitem[{\citenamefont{Perroni et~al.}(2014)\citenamefont{Perroni, Ninno, and
  Cataudella}}]{Perroni2014}
\bibinfo{author}{\bibfnamefont{C.~A.} \bibnamefont{Perroni}},
  \bibinfo{author}{\bibfnamefont{D.}~\bibnamefont{Ninno}}, \bibnamefont{and}
  \bibinfo{author}{\bibfnamefont{V.}~\bibnamefont{Cataudella}},
  \bibinfo{journal}{Phys. Rev. B} \textbf{\bibinfo{volume}{90}},
  \bibinfo{pages}{125421} (\bibinfo{year}{2014}).

\bibitem[{\citenamefont{Zimbovskaya}(2014)}]{Zimbovskaya2014}
\bibinfo{author}{\bibfnamefont{N.~A.} \bibnamefont{Zimbovskaya}},
  \bibinfo{journal}{J. Phys.: Condens. Matter} \textbf{\bibinfo{volume}{26}},
  \bibinfo{pages}{275303} (\bibinfo{year}{2014}).

\bibitem[{\citenamefont{Marcus}(1956)}]{Marcus1956}
\bibinfo{author}{\bibfnamefont{R.~A.} \bibnamefont{Marcus}},
  \bibinfo{journal}{J. Chem. Phys.} \textbf{\bibinfo{volume}{24}},
  \bibinfo{pages}{966} (\bibinfo{year}{1956}).

\bibitem[{\citenamefont{Marcus}(1964)}]{Marcus1964}
\bibinfo{author}{\bibfnamefont{R.~A.} \bibnamefont{Marcus}},
  \bibinfo{journal}{Annu. Rev. Phys. Chem.} \textbf{\bibinfo{volume}{15}},
  \bibinfo{pages}{155} (\bibinfo{year}{1964}).

\bibitem[{\citenamefont{Marcus and Sutin}(1985)}]{Marcus1985}
\bibinfo{author}{\bibfnamefont{R.~A.} \bibnamefont{Marcus}} \bibnamefont{and}
  \bibinfo{author}{\bibfnamefont{N.}~\bibnamefont{Sutin}},
  \bibinfo{journal}{Biochim. Biophys. Acta} \textbf{\bibinfo{volume}{811}},
  \bibinfo{pages}{265 } (\bibinfo{year}{1985}).

\bibitem[{\citenamefont{Marcus}(1993)}]{Marcus1993}
\bibinfo{author}{\bibfnamefont{R.~A.} \bibnamefont{Marcus}},
  \bibinfo{journal}{Rev. Mod. Phys.} \textbf{\bibinfo{volume}{65}},
  \bibinfo{pages}{599} (\bibinfo{year}{1993}).

\bibitem[{\citenamefont{Tachiya}(1993)}]{Tachiya1993}
\bibinfo{author}{\bibfnamefont{M.}~\bibnamefont{Tachiya}}, \bibinfo{journal}{J.
  Phys. Chem.} \textbf{\bibinfo{volume}{97}}, \bibinfo{pages}{5911}
  (\bibinfo{year}{1993}).

\bibitem[{\citenamefont{Nitzan}(2006)}]{Nitzan2006chemical}
\bibinfo{author}{\bibfnamefont{A.}~\bibnamefont{Nitzan}},
  \emph{\bibinfo{title}{Chemical Dynamics in Condensed Phases: Relaxation,
  Transfer and Reactions in Condensed Molecular Systems}}
  (\bibinfo{publisher}{Oxford University Press}, \bibinfo{year}{2006}).

\bibitem[{\citenamefont{Peters}(2015)}]{Peters2015}
\bibinfo{author}{\bibfnamefont{B.}~\bibnamefont{Peters}}, \bibinfo{journal}{J.
  Phys. Chem. B} \textbf{\bibinfo{volume}{119}}, \bibinfo{pages}{6349}
  (\bibinfo{year}{2015}).

\bibitem[{\citenamefont{Zichi et~al.}(1989)\citenamefont{Zichi, Ciccotti,
  Hynes, and Ferrario}}]{Hynes1989}
\bibinfo{author}{\bibfnamefont{D.~A.} \bibnamefont{Zichi}},
  \bibinfo{author}{\bibfnamefont{G.}~\bibnamefont{Ciccotti}},
  \bibinfo{author}{\bibfnamefont{J.~T.} \bibnamefont{Hynes}}, \bibnamefont{and}
  \bibinfo{author}{\bibfnamefont{M.}~\bibnamefont{Ferrario}},
  \bibinfo{journal}{J. Phys. Chem.} \textbf{\bibinfo{volume}{93}},
  \bibinfo{pages}{6261} (\bibinfo{year}{1989}).

\bibitem[{\citenamefont{Tachiya}(1989)}]{Tachiya1989}
\bibinfo{author}{\bibfnamefont{M.}~\bibnamefont{Tachiya}}, \bibinfo{journal}{J.
  Phys. Chem.} \textbf{\bibinfo{volume}{93}}, \bibinfo{pages}{7050}
  (\bibinfo{year}{1989}).

\bibitem[{\citenamefont{Steeger et~al.}(2015)\citenamefont{Steeger, Griesbeck,
  Schmiedel, Holzapfel, Krummenacher, Braunschweig, and Lambert}}]{Steeger2015}
\bibinfo{author}{\bibfnamefont{M.}~\bibnamefont{Steeger}},
  \bibinfo{author}{\bibfnamefont{S.}~\bibnamefont{Griesbeck}},
  \bibinfo{author}{\bibfnamefont{A.}~\bibnamefont{Schmiedel}},
  \bibinfo{author}{\bibfnamefont{M.}~\bibnamefont{Holzapfel}},
  \bibinfo{author}{\bibfnamefont{I.}~\bibnamefont{Krummenacher}},
  \bibinfo{author}{\bibfnamefont{H.}~\bibnamefont{Braunschweig}},
  \bibnamefont{and} \bibinfo{author}{\bibfnamefont{C.}~\bibnamefont{Lambert}},
  \bibinfo{journal}{Phys. Chem. Chem. Phys.} \textbf{\bibinfo{volume}{17}},
  \bibinfo{pages}{11848} (\bibinfo{year}{2015}).

\bibitem[{\citenamefont{Soudackov et~al.}(2011)\citenamefont{Soudackov, Hazra,
  and Hammes-Schiffer}}]{Schiffer2011}
\bibinfo{author}{\bibfnamefont{A.~V.} \bibnamefont{Soudackov}},
  \bibinfo{author}{\bibfnamefont{A.}~\bibnamefont{Hazra}}, \bibnamefont{and}
  \bibinfo{author}{\bibfnamefont{S.}~\bibnamefont{Hammes-Schiffer}},
  \bibinfo{journal}{J. Chem. Phys.} \textbf{\bibinfo{volume}{135}},
  \bibinfo{pages}{144115} (\bibinfo{year}{2011}).

\bibitem[{\citenamefont{Hammes-Schiffer}(2015)}]{Schiffer2015a}
\bibinfo{author}{\bibfnamefont{S.}~\bibnamefont{Hammes-Schiffer}},
  \bibinfo{journal}{J. Am. Chem. Soc.} \textbf{\bibinfo{volume}{137}},
  \bibinfo{pages}{8860} (\bibinfo{year}{2015}).

\bibitem[{\citenamefont{Harshan et~al.}(2015)\citenamefont{Harshan, Yu,
  Soudackov, and Hammes-Schiffer}}]{Schiffer2015b}
\bibinfo{author}{\bibfnamefont{A.~K.} \bibnamefont{Harshan}},
  \bibinfo{author}{\bibfnamefont{T.}~\bibnamefont{Yu}},
  \bibinfo{author}{\bibfnamefont{A.~V.} \bibnamefont{Soudackov}},
  \bibnamefont{and}
  \bibinfo{author}{\bibfnamefont{S.}~\bibnamefont{Hammes-Schiffer}},
  \bibinfo{journal}{J. Am. Chem. Soc.} \textbf{\bibinfo{volume}{137}},
  \bibinfo{pages}{13545} (\bibinfo{year}{2015}).

\bibitem[{\citenamefont{Grunwald}(1985)}]{Grunwald1985}
\bibinfo{author}{\bibfnamefont{E.}~\bibnamefont{Grunwald}},
  \bibinfo{journal}{J. Am. Chem. Soc.} \textbf{\bibinfo{volume}{107}},
  \bibinfo{pages}{125} (\bibinfo{year}{1985}).

\bibitem[{\citenamefont{Guthrie}(1996)}]{Guthrie1996}
\bibinfo{author}{\bibfnamefont{J.~P.} \bibnamefont{Guthrie}},
  \bibinfo{journal}{J. Am. Chem. Soc.} \textbf{\bibinfo{volume}{118}},
  \bibinfo{pages}{12878} (\bibinfo{year}{1996}).

\bibitem[{\citenamefont{Lambert et~al.}(2001)\citenamefont{Lambert, N{\"o}ll,
  and Hampel}}]{Lambert2001}
\bibinfo{author}{\bibfnamefont{C.}~\bibnamefont{Lambert}},
  \bibinfo{author}{\bibfnamefont{G.}~\bibnamefont{N{\"o}ll}}, \bibnamefont{and}
  \bibinfo{author}{\bibfnamefont{F.}~\bibnamefont{Hampel}},
  \bibinfo{journal}{J. Phys. Chem. A} \textbf{\bibinfo{volume}{105}},
  \bibinfo{pages}{7751} (\bibinfo{year}{2001}).

\bibitem[{\citenamefont{Zwickl et~al.}(2008)\citenamefont{Zwickl, Shenvi,
  Schmidt, and Tully}}]{Tully2008}
\bibinfo{author}{\bibfnamefont{J.}~\bibnamefont{Zwickl}},
  \bibinfo{author}{\bibfnamefont{N.}~\bibnamefont{Shenvi}},
  \bibinfo{author}{\bibfnamefont{J.~R.} \bibnamefont{Schmidt}},
  \bibnamefont{and} \bibinfo{author}{\bibfnamefont{J.~C.} \bibnamefont{Tully}},
  \bibinfo{journal}{J. Phys. Chem. A} \textbf{\bibinfo{volume}{112}},
  \bibinfo{pages}{10570} (\bibinfo{year}{2008}).

\bibitem[{\citenamefont{Rubtsov}(2015)}]{Rubtsov2015nature}
\bibinfo{author}{\bibfnamefont{I.~V.} \bibnamefont{Rubtsov}},
  \bibinfo{journal}{Nature Chem.} \textbf{\bibinfo{volume}{7}},
  \bibinfo{pages}{683} (\bibinfo{year}{2015}).

\bibitem[{\citenamefont{Delor et~al.}(2015{\natexlab{a}})\citenamefont{Delor,
  Sazanovich, Towrie, and Weinstein}}]{Delor2015acr}
\bibinfo{author}{\bibfnamefont{M.}~\bibnamefont{Delor}},
  \bibinfo{author}{\bibfnamefont{I.~V.} \bibnamefont{Sazanovich}},
  \bibinfo{author}{\bibfnamefont{M.}~\bibnamefont{Towrie}}, \bibnamefont{and}
  \bibinfo{author}{\bibfnamefont{J.~A.} \bibnamefont{Weinstein}},
  \bibinfo{journal}{Acc. Chem. Res.} \textbf{\bibinfo{volume}{48}},
  \bibinfo{pages}{1131} (\bibinfo{year}{2015}{\natexlab{a}}).

\bibitem[{\citenamefont{Delor et~al.}(2015{\natexlab{b}})\citenamefont{Delor,
  Keane, Scattergood, Sazanovich, Greetham, Towrie, Meijer, and
  Weinstein}}]{Delor2015nature}
\bibinfo{author}{\bibfnamefont{M.}~\bibnamefont{Delor}},
  \bibinfo{author}{\bibfnamefont{T.}~\bibnamefont{Keane}},
  \bibinfo{author}{\bibfnamefont{P.~A.} \bibnamefont{Scattergood}},
  \bibinfo{author}{\bibfnamefont{I.~V.} \bibnamefont{Sazanovich}},
  \bibinfo{author}{\bibfnamefont{G.~M.} \bibnamefont{Greetham}},
  \bibinfo{author}{\bibfnamefont{M.}~\bibnamefont{Towrie}},
  \bibinfo{author}{\bibfnamefont{A.~J.} \bibnamefont{Meijer}},
  \bibnamefont{and} \bibinfo{author}{\bibfnamefont{J.~A.}
  \bibnamefont{Weinstein}}, \bibinfo{journal}{Nature Chem.}
  \textbf{\bibinfo{volume}{7}}, \bibinfo{pages}{689}
  (\bibinfo{year}{2015}{\natexlab{b}}).

\bibitem[{\citenamefont{Hammes-Schiffer and Tully}(1995)}]{Schiffer1995}
\bibinfo{author}{\bibfnamefont{S.}~\bibnamefont{Hammes-Schiffer}}
  \bibnamefont{and} \bibinfo{author}{\bibfnamefont{J.~C.} \bibnamefont{Tully}},
  \bibinfo{journal}{J. Chem. Phys.} \textbf{\bibinfo{volume}{103}},
  \bibinfo{pages}{8528} (\bibinfo{year}{1995}).

\bibitem[{\citenamefont{J\'ohannesson and J\'onsson}(2001)}]{Jonsson2001}
\bibinfo{author}{\bibfnamefont{G.~H.} \bibnamefont{J\'ohannesson}}
  \bibnamefont{and}
  \bibinfo{author}{\bibfnamefont{H.}~\bibnamefont{J\'onsson}},
  \bibinfo{journal}{J. Chem. Phys.} \textbf{\bibinfo{volume}{115}},
  \bibinfo{pages}{9644} (\bibinfo{year}{2001}).

\bibitem[{\citenamefont{Voth et~al.}(1989)\citenamefont{Voth, Chandler, and
  Miller}}]{voth89}
\bibinfo{author}{\bibfnamefont{G.~A.} \bibnamefont{Voth}},
  \bibinfo{author}{\bibfnamefont{D.}~\bibnamefont{Chandler}}, \bibnamefont{and}
  \bibinfo{author}{\bibfnamefont{W.}~\bibnamefont{Miller}},
  \bibinfo{journal}{J. Chem. Phys.} \textbf{\bibinfo{volume}{91}},
  \bibinfo{pages}{7749} (\bibinfo{year}{1989}).

\bibitem[{\citenamefont{Vanden-Eijnden and Tal}(2005)}]{Vanden2005}
\bibinfo{author}{\bibfnamefont{E.}~\bibnamefont{Vanden-Eijnden}}
  \bibnamefont{and} \bibinfo{author}{\bibfnamefont{F.~A.} \bibnamefont{Tal}},
  \bibinfo{journal}{J. Chem. Phys.} \textbf{\bibinfo{volume}{123}},
  \bibinfo{pages}{184103} (\bibinfo{year}{2005}).

\bibitem[{\citenamefont{Hartmann et~al.}(2011)\citenamefont{Hartmann, Latorre,
  and Ciccotti}}]{Hartmann2011deltafunction}
\bibinfo{author}{\bibfnamefont{C.}~\bibnamefont{Hartmann}},
  \bibinfo{author}{\bibfnamefont{J.~C.} \bibnamefont{Latorre}},
  \bibnamefont{and} \bibinfo{author}{\bibfnamefont{G.}~\bibnamefont{Ciccotti}},
  \bibinfo{journal}{Eur. Phys. J. Spec. Top.} \textbf{\bibinfo{volume}{200}},
  \bibinfo{pages}{73} (\bibinfo{year}{2011}).

\bibitem[{\citenamefont{Richardson and Thoss}(2014)}]{Thoss2014}
\bibinfo{author}{\bibfnamefont{J.~O.} \bibnamefont{Richardson}}
  \bibnamefont{and} \bibinfo{author}{\bibfnamefont{M.}~\bibnamefont{Thoss}},
  \bibinfo{journal}{J. Chem. Phys.} \textbf{\bibinfo{volume}{141}},
  \bibinfo{pages}{074106} (\bibinfo{year}{2014}).

\bibitem[{\citenamefont{Newton}(2015)}]{Newton2015}
\bibinfo{author}{\bibfnamefont{M.~D.} \bibnamefont{Newton}},
  \bibinfo{journal}{J. Phys. Chem. B} \textbf{\bibinfo{volume}{119}},
  \bibinfo{pages}{14728} (\bibinfo{year}{2015}).

\bibitem[{\citenamefont{Bartsch et~al.}(2005)\citenamefont{Bartsch, Hernandez,
  and Uzer}}]{dawn05a}
\bibinfo{author}{\bibfnamefont{T.}~\bibnamefont{Bartsch}},
  \bibinfo{author}{\bibfnamefont{R.}~\bibnamefont{Hernandez}},
  \bibnamefont{and} \bibinfo{author}{\bibfnamefont{T.}~\bibnamefont{Uzer}},
  \bibinfo{journal}{Phys. Rev. Lett.} \textbf{\bibinfo{volume}{95}},
  \bibinfo{pages}{058301(1)} (\bibinfo{year}{2005}).

\bibitem[{\citenamefont{Craven et~al.}(2014{\natexlab{a}})\citenamefont{Craven,
  Bartsch, and Hernandez}}]{craven14a}
\bibinfo{author}{\bibfnamefont{G.~T.} \bibnamefont{Craven}},
  \bibinfo{author}{\bibfnamefont{T.}~\bibnamefont{Bartsch}}, \bibnamefont{and}
  \bibinfo{author}{\bibfnamefont{R.}~\bibnamefont{Hernandez}},
  \bibinfo{journal}{Phys. Rev. E} \textbf{\bibinfo{volume}{89}},
  \bibinfo{pages}{040801(R)} (\bibinfo{year}{2014}{\natexlab{a}}).

\bibitem[{\citenamefont{Craven et~al.}(2014{\natexlab{b}})\citenamefont{Craven,
  Bartsch, and Hernandez}}]{craven14c}
\bibinfo{author}{\bibfnamefont{G.~T.} \bibnamefont{Craven}},
  \bibinfo{author}{\bibfnamefont{T.}~\bibnamefont{Bartsch}}, \bibnamefont{and}
  \bibinfo{author}{\bibfnamefont{R.}~\bibnamefont{Hernandez}},
  \bibinfo{journal}{J. Chem. Phys.} \textbf{\bibinfo{volume}{141}},
  \bibinfo{pages}{041106} (\bibinfo{year}{2014}{\natexlab{b}}).

\bibitem[{\citenamefont{Craven et~al.}(2015)\citenamefont{Craven, Bartsch, and
  Hernandez}}]{craven15a}
\bibinfo{author}{\bibfnamefont{G.~T.} \bibnamefont{Craven}},
  \bibinfo{author}{\bibfnamefont{T.}~\bibnamefont{Bartsch}}, \bibnamefont{and}
  \bibinfo{author}{\bibfnamefont{R.}~\bibnamefont{Hernandez}},
  \bibinfo{journal}{J. Chem. Phys.} \textbf{\bibinfo{volume}{142}},
  \bibinfo{pages}{074108} (\bibinfo{year}{2015}).

\bibitem[{\citenamefont{Craven and Hernandez}(2015)}]{craven15c}
\bibinfo{author}{\bibfnamefont{G.~T.} \bibnamefont{Craven}} \bibnamefont{and}
  \bibinfo{author}{\bibfnamefont{R.}~\bibnamefont{Hernandez}},
  \bibinfo{journal}{Phys. Rev. Lett.} \textbf{\bibinfo{volume}{115}},
  \bibinfo{pages}{148301} (\bibinfo{year}{2015}).

\bibitem[{\citenamefont{Tolman}(1920)}]{Tolman1920}
\bibinfo{author}{\bibfnamefont{R.~C.} \bibnamefont{Tolman}},
  \bibinfo{journal}{J. Am. Chem. Soc.} \textbf{\bibinfo{volume}{42}},
  \bibinfo{pages}{2506} (\bibinfo{year}{1920}).

\bibitem[{\citenamefont{Truhlar}(1978)}]{Truhlar1978}
\bibinfo{author}{\bibfnamefont{D.~G.} \bibnamefont{Truhlar}},
  \bibinfo{journal}{J. Chem. Ed.} \textbf{\bibinfo{volume}{55}},
  \bibinfo{pages}{309} (\bibinfo{year}{1978}).

\bibitem[{\citenamefont{Truhlar and Kohen}(2001)}]{Truhlar2001}
\bibinfo{author}{\bibfnamefont{D.~G.} \bibnamefont{Truhlar}} \bibnamefont{and}
  \bibinfo{author}{\bibfnamefont{A.}~\bibnamefont{Kohen}},
  \bibinfo{journal}{Proc. Natl. Acad. Sci.} \textbf{\bibinfo{volume}{98}},
  \bibinfo{pages}{848} (\bibinfo{year}{2001}).

\bibitem[{\citenamefont{Esposito et~al.}(2015)\citenamefont{Esposito, Ochoa,
  and Galperin}}]{Esposito2015}
\bibinfo{author}{\bibfnamefont{M.}~\bibnamefont{Esposito}},
  \bibinfo{author}{\bibfnamefont{M.~A.} \bibnamefont{Ochoa}}, \bibnamefont{and}
  \bibinfo{author}{\bibfnamefont{M.}~\bibnamefont{Galperin}},
  \bibinfo{journal}{Phys. Rev. B} \textbf{\bibinfo{volume}{91}},
  \bibinfo{pages}{115417} (\bibinfo{year}{2015}).

\bibitem[{\citenamefont{Lim et~al.}(2013)\citenamefont{Lim, L\'opez, and
  S\'anchez}}]{Lim2013}
\bibinfo{author}{\bibfnamefont{J.~S.} \bibnamefont{Lim}},
  \bibinfo{author}{\bibfnamefont{R.}~\bibnamefont{L\'opez}}, \bibnamefont{and}
  \bibinfo{author}{\bibfnamefont{D.}~\bibnamefont{S\'anchez}},
  \bibinfo{journal}{Phys. Rev. B} \textbf{\bibinfo{volume}{88}},
  \bibinfo{pages}{201304} (\bibinfo{year}{2013}).

\bibitem[{\citenamefont{Schiff and Nitzan}(2010)}]{Nitzan2010}
\bibinfo{author}{\bibfnamefont{P.~R.} \bibnamefont{Schiff}} \bibnamefont{and}
  \bibinfo{author}{\bibfnamefont{A.}~\bibnamefont{Nitzan}},
  \bibinfo{journal}{Chem. Phys.} \textbf{\bibinfo{volume}{375}},
  \bibinfo{pages}{399 } (\bibinfo{year}{2010}).

\bibitem[{\citenamefont{Chtchelkatchev
  et~al.}(2013)\citenamefont{Chtchelkatchev, Glatz, and
  Beloborodov}}]{Chet2013}
\bibinfo{author}{\bibfnamefont{N.~M.} \bibnamefont{Chtchelkatchev}},
  \bibinfo{author}{\bibfnamefont{A.}~\bibnamefont{Glatz}}, \bibnamefont{and}
  \bibinfo{author}{\bibfnamefont{I.~S.} \bibnamefont{Beloborodov}},
  \bibinfo{journal}{J. Phys.: Condens. Matter} \textbf{\bibinfo{volume}{25}},
  \bibinfo{pages}{185301} (\bibinfo{year}{2013}).

\bibitem[{\citenamefont{Popov and Hernandez}(2007)}]{hern07a}
\bibinfo{author}{\bibfnamefont{A.~V.} \bibnamefont{Popov}} \bibnamefont{and}
  \bibinfo{author}{\bibfnamefont{R.}~\bibnamefont{Hernandez}},
  \bibinfo{journal}{J. Chem. Phys.} \textbf{\bibinfo{volume}{126}},
  \bibinfo{pages}{244506} (\bibinfo{year}{2007}).

\bibitem[{\citenamefont{Uzer et~al.}(2002)\citenamefont{Uzer, Jaff{\'e},
  Palaci{\'a}n, Yanguas, and Wiggins}}]{Uzer02}
\bibinfo{author}{\bibfnamefont{T.}~\bibnamefont{Uzer}},
  \bibinfo{author}{\bibfnamefont{C.}~\bibnamefont{Jaff{\'e}}},
  \bibinfo{author}{\bibfnamefont{J.}~\bibnamefont{Palaci{\'a}n}},
  \bibinfo{author}{\bibfnamefont{P.}~\bibnamefont{Yanguas}}, \bibnamefont{and}
  \bibinfo{author}{\bibfnamefont{S.}~\bibnamefont{Wiggins}},
  \bibinfo{journal}{Nonlinearity} \textbf{\bibinfo{volume}{15}},
  \bibinfo{pages}{957} (\bibinfo{year}{2002}).

\bibitem[{\citenamefont{Thompson}(1998)}]{Thompson1998multi}
\bibinfo{author}{\bibfnamefont{D.~L.} \bibnamefont{Thompson}},
  \emph{\bibinfo{title}{Modern Methods for Multidimensional Dynamics
  Computations in Chemistry}} (\bibinfo{publisher}{World Scientific},
  \bibinfo{year}{1998}).

\bibitem[{\citenamefont{Lykhin et~al.}(2016)\citenamefont{Lykhin, Kaliakin,
  dePolo, Kuzubov, and Varganov}}]{Varganov2016}
\bibinfo{author}{\bibfnamefont{A.~O.} \bibnamefont{Lykhin}},
  \bibinfo{author}{\bibfnamefont{D.~S.} \bibnamefont{Kaliakin}},
  \bibinfo{author}{\bibfnamefont{G.~E.} \bibnamefont{dePolo}},
  \bibinfo{author}{\bibfnamefont{A.~A.} \bibnamefont{Kuzubov}},
  \bibnamefont{and} \bibinfo{author}{\bibfnamefont{S.~A.}
  \bibnamefont{Varganov}}, \bibinfo{journal}{Int. J. Quantum Chem.}
  \textbf{\bibinfo{volume}{116}}, \bibinfo{pages}{750} (\bibinfo{year}{2016}).

\bibitem[{\citenamefont{Menzinger and Wolfgang}(1969)}]{Wolfgang1969}
\bibinfo{author}{\bibfnamefont{M.}~\bibnamefont{Menzinger}} \bibnamefont{and}
  \bibinfo{author}{\bibfnamefont{R.}~\bibnamefont{Wolfgang}},
  \bibinfo{journal}{Angew. Chem., Ind. Ed.} \textbf{\bibinfo{volume}{8}},
  \bibinfo{pages}{438} (\bibinfo{year}{1969}), ISSN \bibinfo{issn}{1521-3773}.

\bibitem[{\citenamefont{Kohen et~al.}(1999)\citenamefont{Kohen, Cannio,
  Bartolucci, and Klinman}}]{Kohen1999}
\bibinfo{author}{\bibfnamefont{A.}~\bibnamefont{Kohen}},
  \bibinfo{author}{\bibfnamefont{R.}~\bibnamefont{Cannio}},
  \bibinfo{author}{\bibfnamefont{S.}~\bibnamefont{Bartolucci}},
  \bibnamefont{and} \bibinfo{author}{\bibfnamefont{J.~P.}
  \bibnamefont{Klinman}}, \bibinfo{journal}{Nature}
  \textbf{\bibinfo{volume}{399}}, \bibinfo{pages}{496} (\bibinfo{year}{1999}).

\end{thebibliography}

\end{document}